\documentclass[useAMS,usenatbib]{mn2e}

\usepackage{graphicx}

\usepackage{natbib}

\usepackage{times}

\usepackage[T1]{fontenc}
\usepackage{ae,aecompl}

\newcommand{\be}{\begin{displaymath}}
\newcommand{\ee}{\end{displaymath}}
\newcommand{\bea}{\begin{eqnarray}}
\newcommand{\eea}{\end{eqnarray}}

\newcommand{\kelv}{\ensuremath{\,\mathrm K}}

\newcommand{\msun}{\ensuremath{\, {\rm M}_\odot}}

\title[Simulations of nova outbursts and nucleosynthesis]{MESA and NuGrid simulations of classical novae: CO and ONe nova nucleosynthesis }

\author[P. A. Denissenkov, J. W. Truran, M. Pignatari et al.]{P. A. 
Denissenkov$^{1,2,3,9}$\thanks{E-mail: pavelden@uvic.ca.}, J. W. Truran$^{3,4}$,  
M. Pignatari$^{5,9}$, R. Trappitsch$^{6,9}$,\newauthor C. Ritter$^{1,7,9}$, F. Herwig$^{1,3,9}$,
U. Battino$^{5,9}$, K. Setoodehnia$^{8,9}$ and B. Paxton$^{10}$\\
$^{1}$Department of Physics \& Astronomy, University of Victoria,
       P.O.~Box 1700, STN CSC, Victoria, B.C., V8W~2Y2, Canada\\
$^{2}$TRIUMF, 4004 Wesbrook Mall, Vancouver, BC V6T~2A3, Canada\\
$^{3}$The Joint Institute for Nuclear Astrophysics, Notre Dame, IN 46556, USA\\
$^{4}$Department of Astronomy and Astrophysics, and Enrico Fermi Institute, University of Chicago, Chicago, IL 60637 USA\\
$^{5}$Department of Physics, University of Basel, Klingelbergstrasse 82, CH-4056 Basel, Switzerland\\
$^{6}$Department of the Geophysical Science and Chicago Center for Cosmochemistry, University of
  Chicago, Chicago, IL 60637, USA\\
$^{7}$Goethe-Universit{\"a}t Frankfurt, Senckenberganlage 31, 60325 Frankfurt am Main, Germany\\
$^{8}$Department of Physics and Astronomy, McMaster University, Hamilton, ON, Canada L8S 4M1\\
$^{9}$NuGrid collaboration\\
$^{10}$Kavli Institute for Theoretical Physics and Department of Physics, Kohn Hall, University of California, Santa Barbara, CA 93106, USA}
\begin{document}

\date{Accepted 2013 December 31. Received 2013 December 31; in original form 2013 December 31}

\pagerange{\pageref{firstpage}--\pageref{lastpage}} \pubyear{2013}

\maketitle

\label{firstpage}

\begin{abstract}
  Classical novae are the result of thermonuclear flashes of hydrogen accreted
  by CO or ONe white dwarfs, leading eventually to the dynamic ejection
  of the surface layers. These are observationally known to be enriched
  in heavy elements, such as C, O and Ne that must originate in layers
  below the H-flash convection zone. Building on our previous work, we
  now present stellar evolution simulations of ONe novae and provide a 
  comprehensive comparison of our models with published ones. Some of our models include
  exponential convective boundary mixing to account for the
  observed enrichment of the nova ejecta even when accreted material has a
  solar abundance distribution. Our models produce maximum temperature
  evolution profiles and nucleosynthesis yields in good agreement with
  models that generate enriched ejecta by assuming that the accreted
  material was pre-mixed. We confirm for ONe novae the result we
  reported previously, i.e.\ we found that $^3$He could be produced {\it
  in situ} in solar-composition envelopes accreted with slow rates
  ($\dot{M} < 10^{-10}\,M_\odot/\mbox{yr}$) by cold ($T_{\rm WD} < 10^7$
  K) CO WDs, and that convection was triggered by $^3$He burning before
  the nova outburst in that case. In addition, we now find that the
  interplay between the $^3$He production and destruction in the
  solar-composition envelope accreted with an intermediate rate, e.g.\
  $\dot{M} = 10^{-10}\,M_\odot/\mbox{yr}$, by the $1.15\,M_\odot$ ONe WD
  with a relatively high initial central temperature, e.g.\ $T_{\rm WD}
  = 15\times 10^6$ K, leads to the formation of a thick radiative buffer
  zone that separates the bottom of the convective envelope from the WD
  surface. We present detailed nucleosynthesis calculations based on the
  post-processing technique, and demonstrate in which way much simpler
  single-zone $T$ and $\rho$ trajectories extracted from the multi-zone stellar
  evolution simulations can be used, in lieu of full multi-zone
  simulations, to analyse the sensitivity of nova abundance predictions
  on nuclear reaction rate uncertainties. Trajectories for both CO and
  ONe nova models for different central temperatures and accretion rates are provided. We
  compare our nova simulations with observations of novae and pre-solar
  grains believed to originate in novae.

\end{abstract}

\begin{keywords}
methods: numerical --- stars: novae --- stars: abundances --- stars: evolution --- stars: interiors
\end{keywords}

\section{Introduction}
\label{sec:intro}

Classical nova explosions are a consequence of thermonuclear
runaways (TNR) occurring in accreted H-rich shells on the
white dwarf (WD) components of close binary systems. The
accumulation of H-rich shells on the surfaces of the
WD components of these systems continues until a
critical pressure $\sim 10^{19} \mathrm{dyne}/\mathrm{cm}^{2}$
is achieved at the base of the
accreted envelope and runaway ensues. An increase in the shell
burning luminosity then follows (at a rate that reflects/defines
the speed class of the nova) to a value that approaches (and for
fast novae can exceed) the Eddington limit. Following a period
of burning at approximately constant bolometric luminosity,
dictated by the core-mass-luminosity relation of \cite{paczynski:70},
the nova ultimately returns to its pre-outburst state.
The occurrence of mass loss from novae through outburst
yields an ejected envelope, the composition of which reflects
both the consequences of the thermonuclear
burning epoch and the occurrence of dredge-up of matter from
the underlying WD into the H-rich envelope.

Our understanding of the nova phenomenon has increased substantially
over the past several decades as a consequence of significant progress
in theoretical modeling of their outbursts, of significantly improved
observational determinations of the characteristics of novae throughout
outburst, and of the acquisition of greatly improved abundance data
concerning both the compositions of the accreted shells and,
particularly, the nebular ejecta of diverse nova systems
\citep[][and references therein]{truran:82,starrfield:89,shara:89,gehrz:98,downen:13}. Multidimensional
hydrodynamic simulations of classical novae are also available
(\citealt{glasner:97,glasner:05,glasner:12,casanova:10,casanova:11,
kercek:98}).

A critical feature of novae seen in outburst is the fact that diverse
nova systems have been found to be characterized by significant
enrichments of many of the elements, such as C, N, O, Ne, Na, Mg, or Al,
relative to H, with respect to solar abundances. It is generally
expected that, for the temperature/density conditions that characterize
the burning shells of most novae (with peak shell temperatures typically less
than $4 \times 10^8 \mathrm{K}$), significant breakout of the CNO
H-burning sequences cannot occur \citep{wiescher:86,yaron:05}. It then follows that the elevated
total abundance of heavy elements between He and Ca, 
as compared to their combined solar mass fraction of less than 2\%, 
arise from dredge-up from the underlying cores, followed by their reshuffling in the H-shell burning.  
C, N, and O are enriched by dredge-up from CO WDs, while the observed
abundance enrichments of C, O, and Ne arise from the outward
mixing or dredge-up of matter from an underlying ONe WD. 

ONe novae are of particular interest for nucleosynthesis because they
develop the highest temperatures during the nova outburst. They also
represent an unexpectedly large fraction of observed novae. Tables of
abundances in nova ejecta were published in both the papers on nova
composition by \cite{truran:86}, \cite{livio:94} and \cite{downen:13} and the review paper
by \cite{gehrz:98}. In the more extended compilation by \cite{gehrz:98}
out of the 20 novae ten are clearly CO novae
(mass fraction ratios of $\mathrm{Ne}/\mathrm{Ne}_\odot < 8$), while seven (three) are clearly
(possibly) ONe novae ($\mathrm{Ne}/\mathrm{Ne}_\odot > 10-15$). However,
the fraction of ONe white dwarfs  in the Galaxy is expected to be much
smaller, based on models of stellar evolution. Carbon ignition happens only
for (super-)Asymptotic Giant Branch (AGB) stars with core masses greater than $\sim
1.05 M_\odot$, and ONe WDs can only form in stars with initial masses
in the corresponding narrow range of $\sim 7-10 M_\odot$
(depending on metallicity and mixing assumptions during H- and He-core
burning phases) that features C ignition but avoids core-collapse
supernova. Although the details of this simulation prediction depend
sensitively on the $^{12}\mathrm{C} + {^{12}\mathrm{C}}$ reaction rate \citep{chen:13} as well
as on the physics details of the C-flame for off-center C ignition
\citep{denissenkov:13c}, present best estimates based on a Salpeter
initial mass function predict that the ratio of CO to ONe WDs is
approximately thirty \citep{livio:94,gil-pons:03}.

\begin{figure*}
\includegraphics[bb = 135 240 500 600]{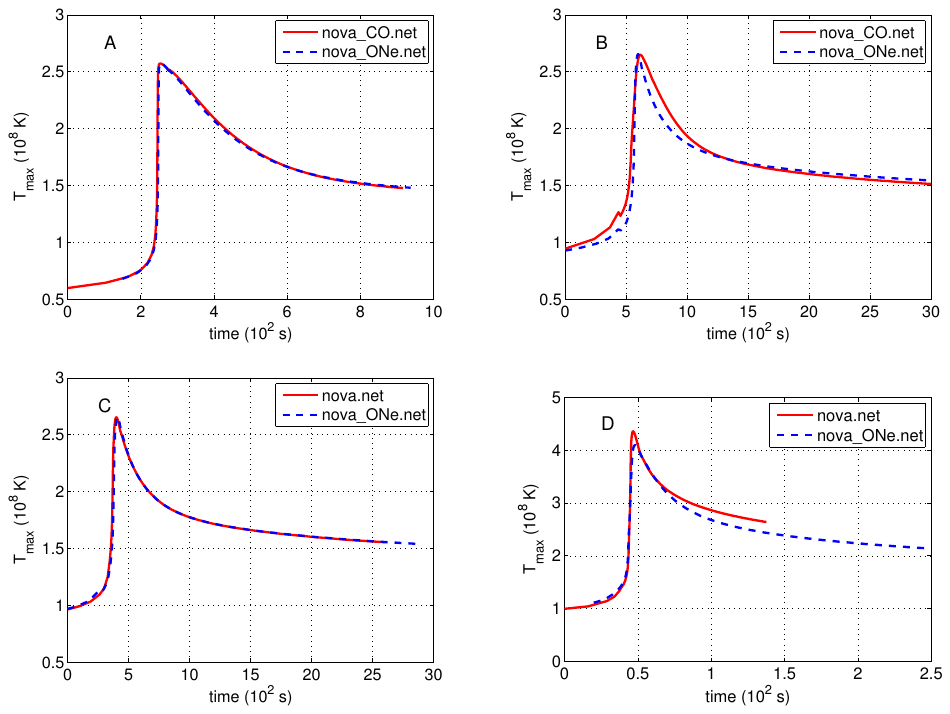}
\caption{Comparison of $T_{\rm max}$-trajectories extracted from our MESA CO and ONe nova simulations 
         that were carried out using different nuclear networks: {\tt nova\_CO.net} that includes only
         33 isotopes from H to $^{26}$Mg coupled by 65 reactions,
         {\tt nova\_ONe.net} that extends the first network to 48 isotopes from H to $^{30}$Si coupled by 120 reactions,
         and the largest network {\tt nova.net} with 77 isotopes from H to $^{40}$Ca and 442 reactions.
         Panel A:  our $1.15\,M_\odot$ CO nova simulations for $T_{\rm WD} = 10$ MK ($\log_{10}L_{\rm WD}/L_\odot = -2.69$;
         for the correspondence between WD's initial central temperature and luminosity,
         $T_{\rm WD}$ and $L_{\rm WD}$, see Table~\ref{tab:tab1})
         and $\dot{M} = 10^{-11}\,M_\odot/\mbox{yr}$. Panels B and C: our simulations of
         $1.15\,M_\odot$ and $1.3\,M_\odot$ ONe novae, respectively, for the same values of $T_\mathrm{WD} = 12$ MK and
         $\dot{M} = 2\times 10^{-10}\,M_\odot/\mbox{yr}$. Panel D: the most extreme case of ONe nova with
         $T_\mathrm{WD} = 7$ MK and $\dot{M} = 10^{-11}\,M_\odot/\mbox{yr}$. 
         All these simulations assume that the WD accretes a mixture of
         equal amounts of its core and solar-composition materials. 
         }
\label{fig:f1}
\end{figure*}

\citet[][hereafter Paper~I]{denissenkov:13} presented 1D simulations of
nova outbursts occurring on CO WDs based on the MESA\footnote{MESA is an
open source code; {\tt http://mesa.sourceforge.net}.} stellar evolution
code \citep{paxton:11,paxton:13}. In the present paper, we complement those results with
MESA models of ONe nova outbursts. In addition, the multi-zone
post-processing code MPPNP, which is a part of the NuGrid research
framework (\citealt{herwig:08}), is used here for computations of
detailed nucleosynthesis in CO and ONe novae. The MESA nova models
provide MPPNP with necessary stellar structure and mixing parameter
profiles. The NuGrid tools also include the single-zone code SPPN that
enables post-processing nucleosynthesis computations along a $(T,\rho)$
trajectory, for example at one Lagrangian coordinate (a mass zone)
inside 1D stellar models, or along the position of the maximum
temperature. We have run SPPN for trajectories tracing the evolution of
the maximum temperature in the nova envelope (usually at the
core-envelope interface) and its corresponding density. We show that,
because of the strong dependence of thermonuclear reaction rates on $T$,
nucleosynthesis yields obtained with the SPPN code for the $T_{\rm
max}$-trajectories are in qualitative agreement with the complete yields
provided by the MPPNP code. The SPPN code runs much faster than MPPNP,
therefore it can be used for a comprehensive numerical analysis of
parameter space, for example, when the relevant reaction rates are
varied within their experimentally constrained limits.

We have prepared Linux shell scripts, template MESA inlist files, and a large
number of initial WD models aiming to combine MESA and NuGrid into an
easy-to-use Nova Framework. 
This new research tool can model nova
outbursts and nucleosynthesis occurring on CO and ONe WDs using
up-to-date input physics and a specified nuclear network for a
relatively large number (up to a few thousands) of mass zones covering
both the WD and its accreted envelope. Nova simulation results can be
analyzed using animations and a variety of plots produced with NuGrid's
Python visualization scripts. One of the main goals of this paper is to
demonstrate Nova Framework's capabilities.

The simulation tools of MESA and NuGrid are described in
Section~\ref{sec:tools}. The MESA models of nova outbursts occurring on
CO and ONe WDs are discussed in Section~\ref{sec:models}. Results of the
post-processing nucleosynthesis computations for CO and ONe novae
obtained with the NuGrid codes MPPNP and SPPN are presented in
Section~\ref{sec:comp}. In Section~\ref{sec:he3mix}, we describe the
effects of $^3$He {\it in situ} production and burning on triggering
convection in the accreted solar-composition envelopes of ONe novae. A
comparison of our results of the nova nucleosynthesis simulations with
available data on nova-related element and isotope abundances from
spectroscopic and presolar grain composition analyses is made in
Section~\ref{sec:observations}. Finally, our main conclusions are drawn
in Section~\ref{sec:concl}.

\begin{figure}
\includegraphics[width = 90mm]{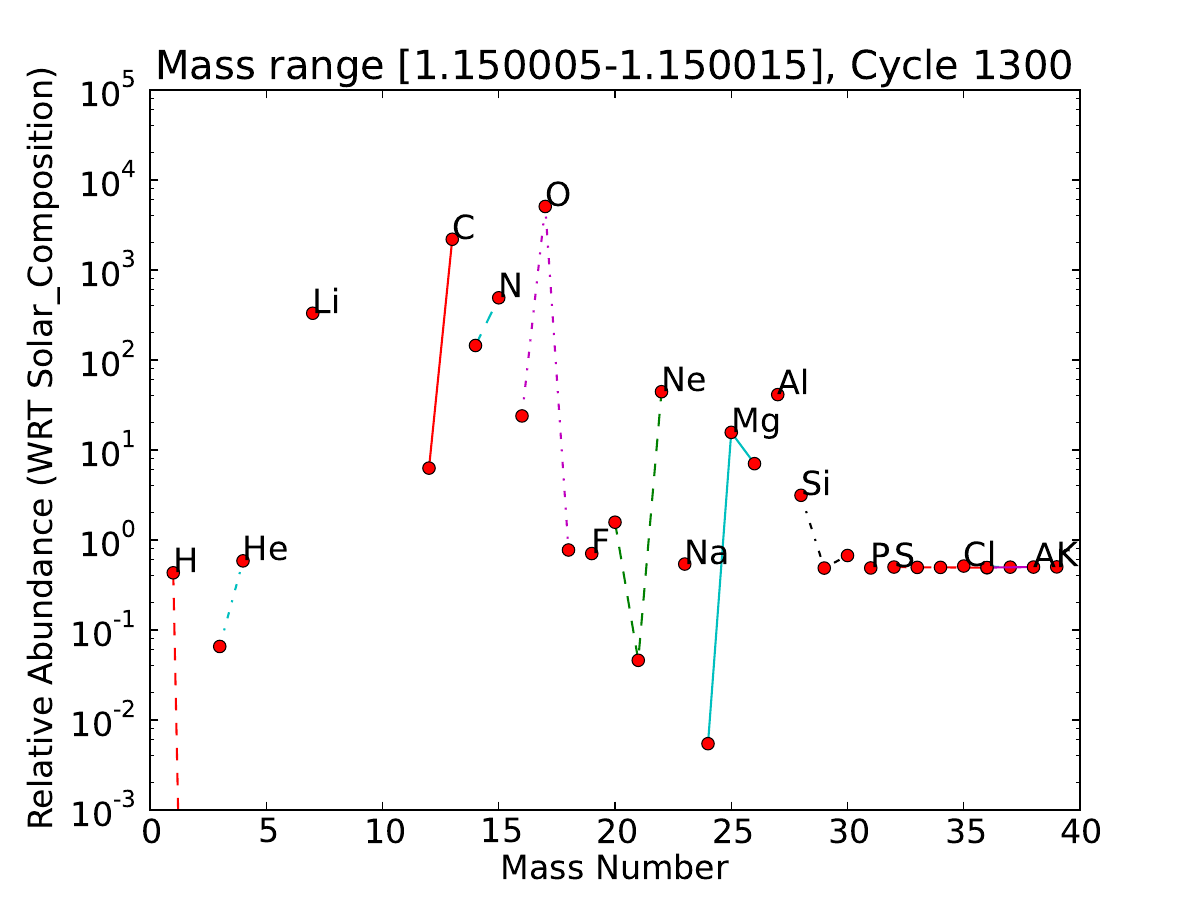}
\caption{Relative (with respect to solar) abundances (mass fractions) of
         stable isotopes calculated with the NuGrid multi-zone post-processing nucleosynthesis code MPPNP, 
         averaged over the indicated mass range (in $M_\odot$) in the expanding envelope, in a final model (its number, 
         called ``Cycle'' in MPPNP, is 1300) of our $1.15\,M_\odot$ CO nova simulations with $T_{\rm WD} = 15$ MK and
         $\dot{M} = 2\times 10^{-10}\,M_\odot/\mbox{yr}$ (compare with Fig.~1 of
         \protect\cite{jose:98} and Fig.~5 in Paper~I). Isotopes of the same elements are connected by lines of
         different colors and styles to better distinguish them visually.
         }
\label{fig:f4}
\end{figure}

\section{Simulation Tools}
\label{sec:tools}

\subsection{Stellar Evolution Calculations -- MESA}
\label{sec:mesa}

MESA is a collection of Fortran-95 {\bf M}odules for {\bf E}xperiments
in {\bf S}tellar {\bf A}strophysics. Its  {\tt star}  module can be
used for 1D stellar evolution simulations of almost any kind (numerous
examples are provided by \cite{paxton:11,paxton:13} and on the MESA users'
website {\tt mesastar.org}). Other MESA modules provide {\tt star} with state-of-the-art numerical
algorithms, e.g. for adaptive mesh refinement and timestep control,
atmospheric boundary conditions, and modern input physics (opacities,
equation of state (EOS), nuclear reaction rates, etc.)

For our MESA simulations of ONe nova outbursts, we use the same EOS and
opacity tables that were used to model the CO novae in Paper~I. The only
important difference is our adoption of longer lists of isotopes and
reactions for the ONe nova models. The adopted nuclear network has to be
as large as necessary in order to account for the energy generation
during a nova TNR, yet as small as possible in order to make
computations not too time consuming. We have
started our MESA nova simulations with the largest network {\tt
nova.net} that includes 77 isotopes from H to $^{40}$Ca coupled
by 442 reactions. This network has a list of isotopes almost identical
to that selected by \cite{weiss:90}, \cite{politano:95}, and
\cite{starrfield:09}. We have gradually reduced the numbers of isotopes
and reactions checking that this does not lead to significant changes in
the time variations of the maximum temperatures and their corresponding densities at the bottoms of accreted
envelopes during nova outbursts that we call ``$T_{\rm max}$-trajectories''. 
As a result, acceptable compromises satisfying the above qualitative criteria have
been found empirically. A list of 33 isotopes coupled by 65 reactions,
including those of the pp chains (the pep reaction, whose importance was
emphasized by \cite{starrfield:09}, has also been added), CNO and NeNa
cycles, was selected for our CO nova models in Paper~I. We will call
this nuclear network {\tt nova\_CO.net}. A similar selection procedure
for the ONe nova models has resulted in a necessity to adopt an extended
nuclear network {\tt nova\_ONe.net} that includes 48 isotopes from H to
$^{30}$Si coupled by 120 reactions. The $T_{\rm max}$-trajectories
extracted from our CO and ONe nova simulations performed for different
sets of nova parameters using the three nuclear networks are compared in
Fig.\,\ref{fig:f1}. From these plots, it is seen that the {\tt nova\_CO.net} and
{\tt nova\_ONe.net} nuclear networks provide an acceptable accuracy
for the simulations of CO and ONe nova outbursts, respectively. Indeed,
the upgrade from the first to the second network does not noticeably change  
the $T_\mathrm{max}$-trajectories in our simulations of the CO nova outbursts, even when
a cold CO WD accretes at a very low rate, which leads to a stronger TNR (panel A). 
Likewise, in all the ONe cases (e.g., panel C), except the most extreme one shown in panel D,
replacing  {\tt nova\_ONe.net} with {\tt nova.net} does not change much their
corresponding $T_\mathrm{max}$-trajectories.

Depending on the values of
initial parameters, nuclear network and assumptions about mixing, an individual MESA nova
simulation on a modern personal computer can take from a few minutes
to a few hours.

By default, MESA uses reaction rates from \cite{caughlan:88} and
\cite{angulo:99}, with preference given to the second source (NACRE). It
includes updates to the NACRE rates for $^{14}$N(p,$\gamma)^{15}$O
(\citealt{imbriani:05}), the triple-$\alpha$ reaction
(\citealt{fynbo:05}), $^{14}$N$(\alpha,\gamma)^{18}$F
(\citealt{gorres:00}), and $^{12}$C$(\alpha,\gamma)^{16}$O
(\citealt{kunz:02}). Although the main nuclear path for a classical nova
is driven by p-capture reactions and $\beta$-decays, the
$\alpha$-reactions are important for establishing the chemical
composition of its underlying WD (MESA has also been used to prepare WD models for our CO and ONe nova simulations, 
see Section~\ref{sec:onemodels}). 

\subsection{Nucleosynthesis Calculations -- NuGrid}
\label{sec:nugrid}

The  nucleosynthesis calculations were performed with the
post-processing code PPN \citep[either for single-zone or multi-zone
post-processing, ][]{herwig:08,Pignatari:2012dw}, where the input data of
stellar structure (here from MESA models) are processed using a
dynamically updated list of nuclear species and reaction rates. The
network can include more than 5000 species, between H and Bi, and more
than 50000 nuclear reactions. The self-adjusting dynamical network has
at any point in the simulation (and in each layer for multi-zone
simulations) a suitable size, based on the strength of
nucleosynthesis fluxes, $\delta Y_{\rm i}/\delta t_{\rm j}$, that show the variation rates of the
abundances $Y_{\rm i} = X_{\rm i}/A_{\rm i}$ due to the reactions $j$.
Rates are taken from different data sources: the
European NACRE compilation \citep[][]{angulo:99} and \cite{iliadis:01},
or more recent, if available
\citep[e.g.,][]{fynbo:05,kunz:02,imbriani:05}, and the JINA Reaclib v1.1
library \citep[][]{cyburt:10}. 
For neutron capture rates by stable
nuclides and a selection of relevant unstable isotopes, NuGrid uses the
Kadonis compilation ({\tt http://www.kadonis.org}), although neutron captures are not important
for the present work. For weak reaction
rates, we choose between \cite{fuller:85}, \cite{oda:94},
\cite{langanke:00} and \cite{goriely:99}, according to the mass region.

For the complete post-processing of the full MESA nova models, we use
the multi-zone parallel frame of PPN (MPPNP). For single
$(T,\rho)$ trajectories, the post-processing is done with the single-zone version SPPN.
Both variants use the same nuclear physics library and the same package
to solve the nucleosynthesis equations.

\section{MESA Models of Nova Outbursts}
\label{sec:models}

In Paper~I, we presented  MESA models of CO nova outbursts for a number of
initial CO WD masses and temperatures (luminosities). For completeness,
we include some of the CO nova outburst models here as well, and then present the new
ONe nova models. The nucleosynthesis in both model families is then
analysed using the NuGrid tools.

\subsection{CO Novae}
\label{sec:comodels}
A few CO nova models are included in the present discussion as a link to
Paper~I and also as an illustration of application of the NuGrid MPPNP
code for post-processing of CO nova models. All the results relevant to
the CO nova models are presented in Figs.~\ref{fig:f1}A, \ref{fig:f4}, \ref{fig:f2},
and at the beginning of Table~\ref{tab:tab1}. 
Fig.~\ref{fig:f1}A compares two $T_{\rm max}$-trajectories extracted from
our MESA simulations of nova outbursts occurring on a CO WD with the
same mass, $M_{\rm WD} = 1.15\,M_\odot$, and initial central
temperature, $T_{\rm WD} = 10$ MK ($\log_{10}L_{\rm WD}/L_\odot =
-2.69$; for the correspondence between WD's initial central temperature
and luminosity, $T_{\rm WD}$ and $L_{\rm WD}$, see
Table~\ref{tab:tab1}), and for the same accretion rate, $\dot{M} =
10^{-11}\,M_\odot/\mbox{yr}$. The only difference between the models is
that the one has been computed using the small nuclear network {\tt
nova\_CO.net} that includes 33 isotopes from H to $^{26}$Mg coupled by 65
reactions, while the other with the extended network of 48 isotopes and
120 reactions, {\tt nova\_ONe.net}. The good agreement between the two
trajectories, even for a relatively low $T_{\rm WD}$ and slow accretion
rate that lead to a stronger outburst, confirms our choice of {\tt
nova\_CO.net} as a sufficient nuclear network for the MESA CO nova models in
Paper~I. Fig.~\ref{fig:f4} shows the results of the post-processing nucleosynthesis simulations
carried out for one of the CO nova models considered in Paper~I.
Our final abundances of the stable isotopes divided by their corresponding
solar abundances change with the increasing mass number in a fashion closely resembling
that seen in Fig.\,1 of \cite{jose:98} for a CO nova model with similar parameters as well as that
produced by the MESA code alone (see Fig.~5 in Paper~I).
However, when our nova yields are compared with those from the literature
for individual isotopes, the differences can reach an order of magnitude
for some isotopes, or even a larger factor for a few other isotopes.
A similar conclusion can also be made when one compares on an isotope by isotope basis
the nova yields published by \cite{jose:98} and \cite{starrfield:09}.
This is probably caused by differences in many ingredients, such as initial abundances,
reaction rates, computer code details etc., that are used to model nova outbursts.

\begin{table*}
\caption{Parameters of Computed Nova Models}
\label{tab:tab1}
\begin{tabular}{ccccccccc}
\hline
WD & $M_{\rm WD}[\msun]$ & $T_{\rm WD}[10^6\kelv]$ & $\lg L_{WD}[\mathrm{L_\odot}]$ &
         $M_\mathrm{acc}[10^{-5}\msun]$ & $\lg L_\mathrm{H}[\mathrm{L_\odot}]$ &
         $T_\mathrm{max}[10^6\kelv]$ & Nucl. Network & Init. Comp. \\
\hline
CO & 1.0 & 12 & -2.55 & 5.2 & 10.88 & 200 & nova\_CO & MESA \\
CO & 1.0 & 20 & -1.99 & 2.1 & 9.59 & 163 & nova\_CO & MESA  \\
CO$^{a,b}$ & 1.15 & 10 & -2.69 & 4.0 & 11.84 & 257 & {\bf nova\_CO}, nova\_ONe & MESA \\
CO & 1.15 & 12 & -2.50 & 2.8 & 11.27 & 236 & nova\_CO & MESA \\
CO$^{c}$ & 1.15 & 12 & -2.50 & 9.7 & 11.28 & 243 & nova\_CO & MESA \\
CO & 1.15 & 15 & -2.25 & 1.6 & 10.45 & 208 & nova\_CO & MESA \\
CO & 1.15 & 20 & -1.94 & 0.94 & 9.67 & 185 & nova\_CO & MESA \\
ONe$^{b}$ & 1.15 & 12 & -2.54 & 4.0 & 10.24 & 262 & {\bf nova\_CO}, nova\_ONe & MESA \\
ONe$^{b}$ & 1.15 & 12 & -2.54 & 4.0 & 10.14 & 263 & {\bf nova\_ONe}, nova & MESA \\
ONe$^{c}$ & 1.15 & 12 & -2.54 & 15.0 & 9.89 & 245 & nova\_ONe & MESA \\
ONe & 1.15 & 20 & -2.03 & 3.0 & 9.88 & 245 & nova\_ONe & MESA \\
ONe & 1.15 & 20 & -2.03 & 2.0 & 9.46 & 219 & nova\_ONe & Barcelona \\
ONe$^{a}$ & 1.3 & 7 & -3.05 & 3.0 & 10.82 & 408 & {\bf nova\_ONe}, nova & MESA \\
ONe & 1.3 & 12 & -2.52 & 1.7 & 10.23 & 357 & nova\_ONe & Politano \\
ONe & 1.3 & 12 & -2.52 & 1.5 & 10.60 & 344 & nova\_ONe & Barcelona \\
ONe & 1.3 & 15 & -2.29 & 0.95 & 9.90 & 313 & nova\_ONe & Barcelona \\
ONe & 1.3 & 20 & -1.98 & 1.0 & 9.86 & 318 & {\bf nova\_ONe}, nova & MESA \\
ONe & 1.3 & 20 & -1.98 & 0.52 & 9.26 & 267 & nova\_ONe & Barcelona \\
ONe & 1.3 & 20 & -1.98 & 0.53 & 9.29 & 266 & nova (NACRE) & Barcelona \\
ONe & 1.3 & 20 & -1.98 & 0.52 & 9.37 & 265 & nova (JINA) & Barcelona \\
\hline
\end{tabular}

\medskip   

$^{a}$This model uses the accretion rate $\dot{M} = 10^{-11}\,M_\odot/\mbox{yr}$, while all other
$\dot{M} = 2\times 10^{-10}\,M_\odot/\mbox{yr}$.

$^{b}$For simulations done for two different nuclear networks, e.g. {\tt nova\_ONe.net} and {\tt nova.net},
only the results for the first case are shown (for the network highlighted with a bold font).

$^{c}$This simulation includes CBM with $f=0.004$, like in Paper~I, while all other assume that the WD accretes
a mixture of equal amounts of its core and solar-composition materials. For this case, $M_\mathrm{acc}$ gives the mass of
H-rich envelope that includes WD's outer layers penetrated by the CBM.

\end{table*}

\subsection{ONe Novae}
\label{sec:onemodels}

According to \cite{gil-pons:03}, the frequency of occurrence of galactic
ONe novae in close binary systems is expected to be 30\% to 40\%,
depending on model assumptions. Without convective overshooting of
H-core burning, the range of initial masses of the stars that end their
evolution as ONe WDs with the masses between $1.1 M_\odot$ and $1.33
M_\odot$ is estimated by the same authors to be between $9.3 M_\odot$
and $12 M_\odot$.

Two ONe WD models with the masses $M_{\rm WD} = 1.15\,M_\odot$ and
$1.3\,M_\odot$ have been prepared by us with MESA for a range of initial central
temperatures and luminosities using the same ``stellar engineering''
procedure with which we created the CO WD models in Paper~I
(Table~\ref{tab:tab1}, note that the $L_\mathrm{WD}(T_\mathrm{WD})$
relation is almost universal for the CO and ONe WDs for
$M_\mathrm{WD}\geq 1.0 M_\odot$). Likewise, we have artificially removed
the He- and C-rich buffer zones from the surfaces of ONe WD models and
used the naked ONe WDs with thin ($\sim$\,$10^{-7}\,M_\odot$) H-rich
envelopes as the initial models for ONe nova simulations. The ONe WD
models were created using the same isotopes as in {\tt nova.net},
while the reaction list was extended to take into account He and C
burning.

\begin{figure}
\includegraphics[scale = 0.5, bb = 50 220 500 560]{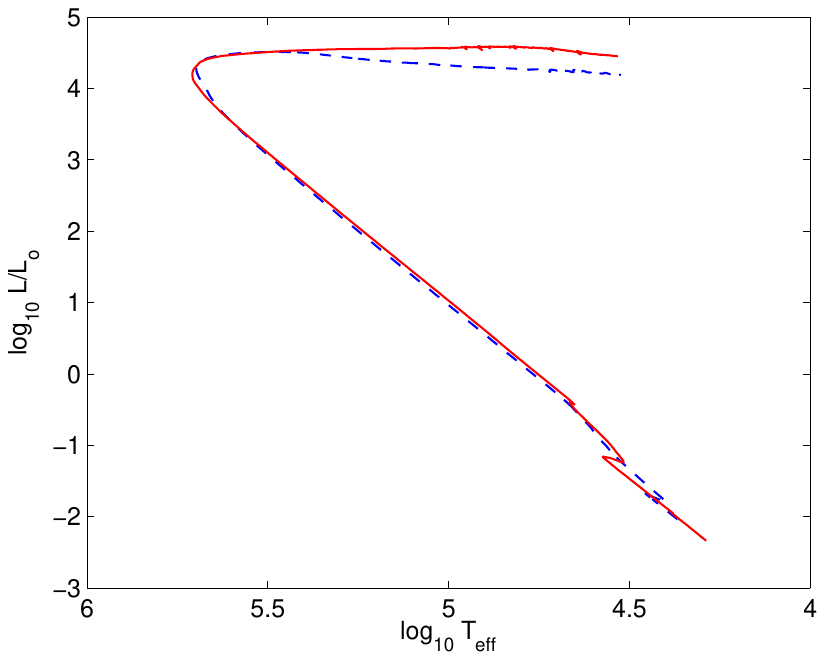}
\caption{Evolutionary tracks for pre-mixed CO (solid red curve) and ONe (dashed blue curve) nova models 
         computed with the MESA code for the same initial parameters: $M_\mathrm{WD} = 1.15 M_\odot$, $T_\mathrm{WD} = 12$ MK, and 
         $\dot{M} = 2\times 10^{-10} M_\odot/\mathrm{yr}$. 
        }
\label{fig:f2}
\end{figure}

We have not included any prescription for mass loss in the MESA ONe nova
simulations, yet we have followed the hydrodynamic expansion of the nova
envelopes until their surface radii exceeded a few solar radii. This
should guarantee that the expanding envelope has crossed the WD's
Roche-lobe radius for a typical binary system hosting a classical nova
(e.g., Paper I). The ONe nova evolutionary tracks are not much different
from our CO nova tracks discussed in Paper I (e.g., Fig.~\ref{fig:f2}).
A more detailed discussion of nova model light curves is beyond the
scope of the present study, which is mainly focused on the nova
outburst phase and nucleosynthesis.

Like for the CO nova simulations in Paper I, here we consider three
types of ONe nova models. A few models are  of the first type: they neglect
boundary mixing (adopting a strict Schwarzschild condition to determine
the convective boundary) and assume that the WD accretes solar-composition
material (e.g., the models with the {\em in situ} $^3$He production
that are discussed in Section~\ref{sec:he3mix}). 
For solar composition we use the elemental abundances of \cite{grevesse:93}
with the isotopic abundance ratios from \cite{lodders:03}. These models do
neither show the fast rise time  nor the enhancements in heavy elements
observed for novae. 

A realistic nova model should assume that the accreted material has
solar composition (for solar metallicity) and that it is mixed with the
WD's material before or during its thermonuclear runaway. Recent two-
and three-dimensional nuclear-hydrodynamic simulations of a nova
outburst have shown that a possible mechanism of this mixing are the
hydrodynamic instabilities and shear-flow turbulence induced by steep
horizontal velocity gradients at the bottom of the convection zone
triggered by the runaway \citep{casanova:10,casanova:11}.
These hydrodynamic processes associated with the convective boundary
lead to convective boundary mixing (CBM) at the base of the accreted
envelope into the outer layers of the WD. As a result, CO-rich (or
ONe-rich) material is dredged up during the runaway. In our 1D MESA
simulations of nova outbursts, we use a simple CBM model that treats the
time-dependent mixing as a diffusion process. The model approximates the
rate of mixing by an exponentially decreasing function of a distance
from the formal convective boundary \citep{freytag:96,herwig:97}
\bea D_{\rm CBM} = D_0\exp\left(-\frac{2|r-r_0|}{fH_P}\right),
\label{eq:DCBM}
\eea 
where $H_P$ is the pressure scale height, and $D_0$ is a diffusion
coefficient, calculated using a mixing-length theory, that describes
convective mixing at the radius $r_0$ close to the boundary. In this
model $f$ is a free parameter, for which we use the same value $f =
f_\mathrm{nova} = 0.004$ that we used in our CO nova simulations in Paper~I,
where it led to the heavy-element enrichment of
nova envelope comparable to those found in both multi-dimensional hydrodynamic simulations
and in the spectroscopic measurements of heavy-element abundances in nova ejecta.
We have varied $f_\mathrm{nova}$ to estimate the sensitivity of our results
to its value. It turns out that very similar nova nucleosynthesis yields are obtained
when we increase $f_\mathrm{nova}$ from 0.004 to 0.006, while
outside of this range the CBM dredges up either too little or too much
WD's material to match the observational data.
The second type of our nova models takes into account the CBM
(e.g., the CBM models in Figs.\,\ref{fig:f11} and \ref{fig:f12}).
These models do match observed rise times and abundances. 

Finally, most of our models (Table \ref{tab:tab1}) are of type three in
which  CBM is neglected, like in the first-type models. However,  the
effect of CBM is artificially obtained by assuming that the WD accretes
a mixture in which solar-composition material has already been blended
with an equal amount of WD material.

\begin{figure}
\includegraphics[scale = 0.75, bb = 135 220 500 700]{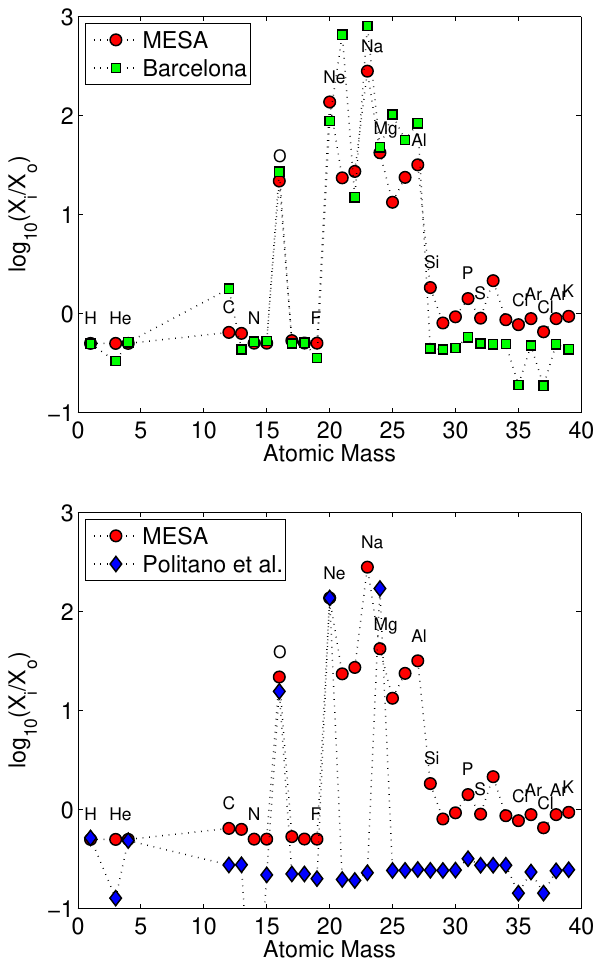}
\caption{Initial abundances (solar-scaled mass fractions) of isotopes in the pre-mixed material accreted by our
         $1.3\,M_\odot$ ONe WD model (red circles) are compared with those used in similar nova models by the Barcelona
         group (\citealt{jose:99,jose:01}) (green squares in the upper panel) and by both
         \protect\cite{politano:95} and \protect\cite{starrfield:09} (blue diamonds in the lower panel). In all the cases,
         it is assumed that the WD accretes a mixture consisting of equal amounts of its core
         and solar-composition materials. Element symbols point to their corresponding most abundant isotopes in the Solar
         system ($^{12}$C, $^{14}$N, $^{16}$O, $^{20}$Ne, $^{23}$Na, $^{24}$Mg, $^{27}$Al, etc.)
         }
\label{fig:f3}
\end{figure}

Such pre-mixed models are commonly adopted in 1D nova simulations. The
fraction of WD's material in the mixture can vary from 25\% up to 75\%
to match observed values (e.g., \citealt{jose:98}). To prepare the
initial pre-mixed abundances, we have used the isotope abundances from
the outermost layers of our naked ONe WD models. Our initial pre-mixed
abundances for the $1.3 M_\odot$ ONe nova model (of the third type) are
compared with the 50\% pre-mixed abundances that we call the Barcelona
and Politano compositions in Fig.~\ref{fig:f3}. The Barcelona data are
taken from \cite{jose:99,jose:01}. They contain material from a $1.35
M_\odot$ ONe WD model. The Politano data are from \cite{politano:95}.
They were derived from the C-burning nucleosynthesis calculations of
\cite{arnett:69}. Our initial abundances agree quite well with those
from the Barcelona composition (the upper panel), especially for
$^{16}$O, $^{20}$Ne, and $^{24}$Mg. As for the Politano composition (the
lower panel of Fig.~\ref{fig:f3}), it has similar to ours $^{16}$O and
$^{20}$Ne abundances, but much higher initial $^{24}$Mg abundance (10\% in mass fraction
versus 3\% in the Barcelona mixture). The
last difference is important for ONe nova simulations because $^{24}$Mg
can compete with $^{12}$C in igniting the accreted H-rich material in
the range of temperatures relevant to nova outbursts
(\citealt{glasner:12}). Our initial $^{12}$C abundance has a value
intermediate between the 6\% and 0.9\% mass fractions in
the Barcelona and Politano mixtures. The more recent massive AGB models \citep{ritossa:96}, including ours,
show that the $^{24}$Mg abundance in ONe WDs is much smaller than
the one assumed by \cite{politano:95}.

\begin{figure}
\includegraphics[scale = 0.75, bb = 135 160 500 650]{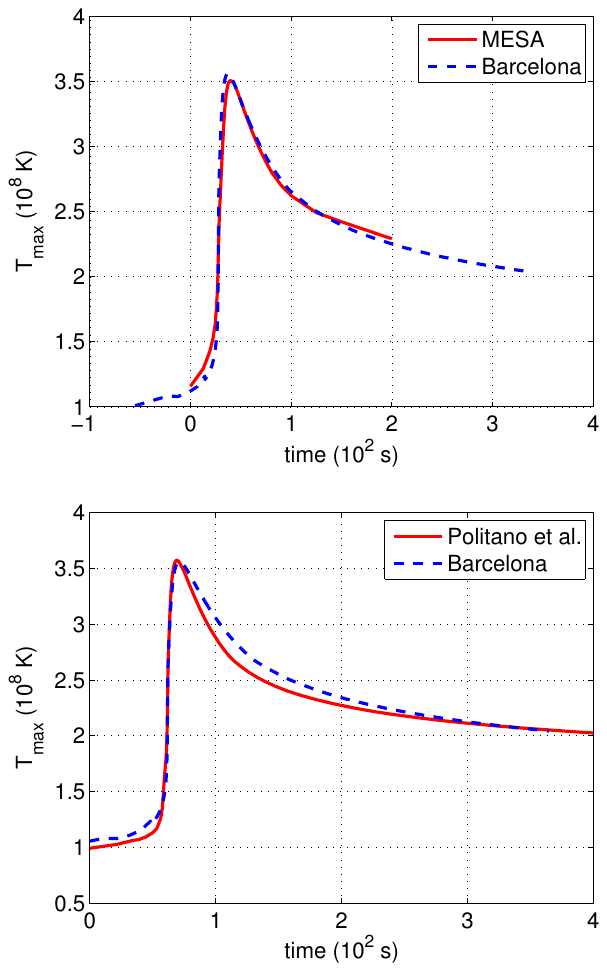}
\caption{$T_\mathrm{max}$-trajectories from our simulations of $1.3\,M_\odot$ ONe novae carried out
         for the same values of $T_\mathrm{WD} = 12$ MK and $\dot{M} = 2\times 10^{-10}\,M_\odot/\mbox{yr}$, but
         for our considered three different initial compositions of the accreted material from Fig.~\ref{fig:f3}.
         }
\label{fig:f3add}
\end{figure}

In Fig.\,\ref{fig:f1}, we compare $T_\mathrm{max}$-trajectories
extracted from our $1.15 M_\odot$ (panel B) and $1.3 M_\odot$
(panel C) ONe nova simulations for $T_\mathrm{WD} = 12$ MK with
pre-mixed material, in which isotope abundances from WD models of the
corresponding masses were used, accreted with the rate $\dot{M} =
2\times 10^{-10} M_\odot/\mathrm{yr}$. For either CO or ONe WD, the main
properties of its nova outburst, such as total accreted
($M_\mathrm{acc}$) and ejected masses, peak temperature
$T_\mathrm{max}$, maximum H-burning ($L_\mathrm{H}$) and total
luminosities, envelope expansion velocity, and chemical composition of
the ejecta, depend mainly on the following three parameters: the WD mass
$M_{\rm WD}$, its central temperature $T_{\rm WD}$ (or luminosity
$L_\mathrm{WD}$), and the accretion rate $\dot{M}$ (e.g.,
\citealt{prialnik:95,townsley:04,jose:07a}). The results also depend on
a choice of nuclear network (lists of isotopes and reactions), initial
abundances in the accreted material, reaction rates and other input
physics (EOS, opacities), and computer code. For the present work, some
of the initial and calculated nova model parameters are listed in
Table~\ref{tab:tab1}. Here, we have only considered material (and
mixtures thereof) of solar metallicity. In general, our pre-mixed
ONe nova models computed with the Barcelona initial composition have
systematically lower $M_\mathrm{acc}$ and $T_\mathrm {max}$ values, as
compared to their counterparts computed with both MESA and Politano
compositions. This is probably caused by the higher initial abundance of
$^{12}$C in the Barcelona mixture (the upper panel of Fig.~\ref{fig:f3})
that turns on the nova ignition a bit earlier.

To categorize our nova simulations carried out for the different sets of
initial parameters, we use $T_\mathrm{max}$-trajectories (e.g.,
Fig.~\ref{fig:f1}). They have two characteristics that determine the
resulting nova speed class and nucleosynthesis: a peak value of the
$T_\mathrm{max}$ curve and its width near the peak, the latter being
proportional to a timescale of nova outburst. We have already used
$T_\mathrm{max}$-trajectories to support our choice of {\tt nova\_CO.net} as
a sufficient nuclear network for MESA CO nova simulations
(Fig.~\ref{fig:f1}A). Similarly, panels B and C of
Fig.~\ref{fig:f1} show that our extended nuclear network {\tt nova\_ONe.net} is acceptable
for MESA ONe nova simulations, except for the most extreme cases, like
that shown in panel D, because its further extension to {\tt
nova.net} does not change the trajectory (panel C), while using
of {\tt nova\_CO.net} results in a slower nova outburst (panel B). The lower panel of Fig.~\ref{fig:f3add}
demonstrates that changing the initial composition in the accreted
envelope from the Barcelona to Politano one produces differences in the
$T_\mathrm{max}$-trajectories comparable to those caused by the
replacing of {\tt nova\_CO.net} by {\tt nova\_ONe.net} in Fig.~\ref{fig:f1}B. The small
difference between the curves in the upper panel of Fig.~\ref{fig:f3add} confirms the conclusion derived
from Fig.~\ref{fig:f3} that our pre-mixed initial composition is closer
to the Barcelona one.

The $T_\mathrm{max}$-trajectories and their corresponding density
profiles can be used by the NuGrid SPPN code for post-processing
computations that trace nucleosynthesis for the specified variations of
$T$ and $\rho$ with time. A shortcoming of such computations is that
they do not take into account abundance changes occurring in other parts
of the nova envelope that are connected by convective mixing. Much more
consistent post-processing nucleosynthesis simulations can be done with
the NuGrid MPPNP code that uses detailed radial profiles of $T$, $\rho$
and the diffusion coefficient from the stellar evolution simulation to
account in the same way for time-dependent mixing. Our Nova Framework
uses a MESA option that allows to generate such input stellar models for
MPPNP, during computations of nova evolution. Still,
given that thermonuclear reaction rates strongly depend on $T$, even
one-zone SPPN simulations should probably produce results that are, at
least qualitatively, compatible with those obtained in full MPPNP
simulations, as discussed in more detail in the
next two sections.

\section{Post-Processing Computations of Nova Nucleosyntheis}
\label{sec:comp}

\subsection{Multi-Zone Computations}
\label{sec:mppnp}
\begin{figure*}
\includegraphics[scale = 0.7, bb = 10 10 600 410]{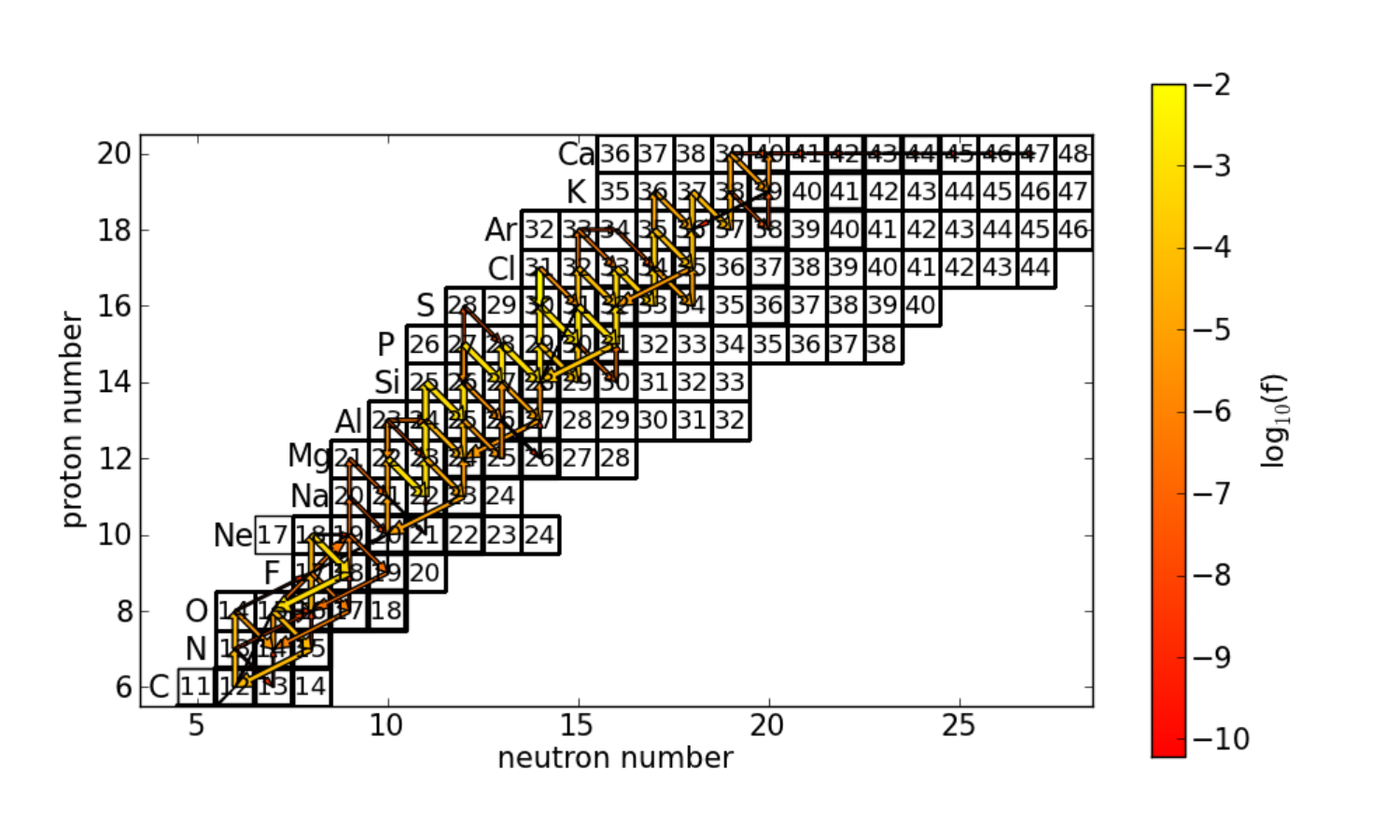}
\caption{The network of 147 isotopes (H, He, Li, Be, and B isotopes are not plotted) used in our
         post-processing nucleosynthesis calculations.
         Reaction fluxes (arrows) correspond to our SPPN simulation for the dashed blue trajectory
         from upper panel of Fig.~\ref{fig:f1}D to a moment when the temperature reaches its peak value of $T_\mathrm{max} = 408$ MK
         (Table~\ref{tab:tab1}). The nucleosynthesis flux, $f = \delta Y_{\rm i}/\delta t_{\rm j}$, 
         shows the variation rate of the
         abundance $Y_{\rm i} = X_{\rm i}/A_{\rm i}$ due to the reaction $j$.
         The arrow width and color correspond to the flux strength. Heavy-lined boxes correspond to the stable isotopes.
         }
\label{fig:f5}
\end{figure*}

MESA simulations use a limited nuclear network. The limit is
imposed by the number of isotopes and reactions needed to accurately
simulate the nuclear energy generation rate which, at the same time,
should not be too large to make MESA computations prohibitively slow.
The NuGrid multi-zone code MPPNP allows to study nucleosynthesis in
stars more accurately than MESA because it can include nuclear networks
with more than 5000 isotopes coupled by more than 50000 reactions (see
\S~\ref{sec:nugrid}, and references therein). This is possible because
MPPNP solves for mixing and burning in an operator split while MESA,
which also has to solve for the structure and energy equations benefits
from the much improved convergence properties of a fully coupled
operator approach\footnote{This means that MESA solves equations of nuclear kinetics
simultaneously with diffusion terms added to them, while MPPNP first solves kinetic equations without
diffusion terms and only after that it mixes resulting abundances, making several iterations for this
two-step procedure.}. MPPNP works in a post-processing mode using stellar
structure models (cycles) prepared by MESA as an input. For nova
nucleosynthesis simulations, we have identified a list of 147 isotopes
coupled by more than 1700 reactions. This nuclear network
includes all the species relevant for nova nucleosynthesis
(Fig.~\ref{fig:f5}). As an example, we show the magnitudes of the main
reaction fluxes for the trajectory with the largest temperature
($T_\mathrm{max}$ = 408 MK) in Fig.~\ref{fig:f5}.
A test run with the full list of nearly 5000 isotopes did not noticeably change
our final nova abundances obtained with the network consisting of 147 isotopes.

\begin{figure}
\includegraphics[scale = 0.75, bb = 135 220 500 700]{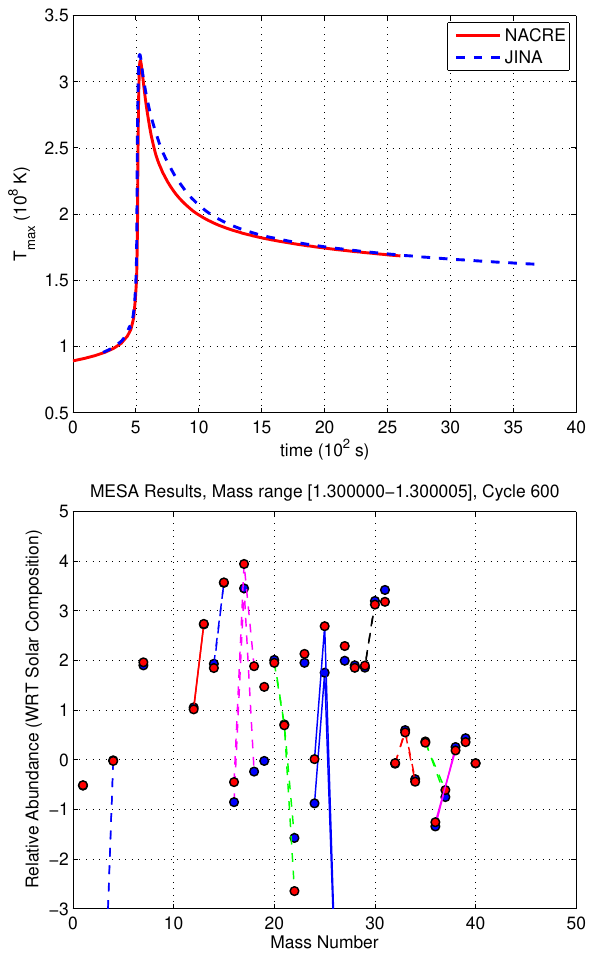}
\caption{Upper panel: $T_{\rm max}$-trajectories from two $1.3\,M_\odot$ pre-mixed ONe nova models with $T_{\rm WD}=20$ MK and
         $\dot{M}=2\times 10^{-10}\,M_\odot/\mbox{yr}$ computed using the same largest nuclear network
         ({\tt nova.net}) but different
         MESA reaction rate data, NACRE and JINA Reaclib. Lower panel: solar-scaled final abundances for these two models extracted
         from their MESA output files. Compare the red symbols from the lower panels in this figure and
         Fig.~\ref{fig:f7}, the latter
         having been obtained in the post-processing nucleosynthesis computations with NuGrid's MPPNP code.
         }
\label{fig:f6}
\end{figure}

As a test of our MESA nova simulations, we have tried the MESA
option for choosing the reaction rates that gives preference to the
JINA Reaclib v1.1 \citep{cyburt:10}, which includes the reaction
rates evaluated  by \cite{iliadis:10}. 
In spite of the presence of some differences in our nova $T_{\rm max}$-trajectories
between the two options that preferentially choose the older (NACRE) and newer reaction rates (e.g., see the upper panel in
Fig.~\ref{fig:f6} and compare the last two rows in Table~\ref{tab:tab1}), we have decided
to use the default MESA option, as we did in Paper~I. The switching between the two options is not likely
to have a large effect on the nova hydrodynamic simulations but may impact nova nucleosynthesis.
However, the latter is computed as post-processing of MESA nova models by the NuGrid code MPPNP that uses
the most recent available reaction rates, including those from the JINA Reaclib.

In Fig.~\ref{fig:f6}, upper panel, we show the
$T_\mathrm{max}$-trajectories from two $1.3 M_\odot$ ONe nova models
with $T_\mathrm{WD} = 20$ MK that accreted the pre-mixed material with
the rate $\dot{M} = 2\times 10^{-10} M_\odot/\mathrm{yr}$ for two different
choices of reaction rate sources available in MESA, NACRE and JINA
Reaclib. The two trajectories do not differ much from one another.
In particular, their corresponding nuclear energy generation budget
differs only by 16\%.

The lower panel of Fig.~\ref{fig:f6} shows the production factors (mass
fractions normalized to the solar composition) for stable isotopes whose
abundances were averaged over the mass of the expanding nova envelope
for the two discussed nova models.
The nucleosynthesis yields are comparable in the two cases, with the
largest differences for $^{22}$Ne and Mg isotopes.

In Fig.~\ref{fig:f4}, the results of our MPPNP
simulations are shown for one of the CO nova models from Paper~I for
which a similar plot, but with data extracted from MESA simulations, was
displayed in Fig.~5 of Paper~I. 
Fig.~1 of \cite{jose:98} presents, in the same form, the nucleosynthesis yields
for a CO nova model with the similar parameters.
The data displayed in Fig.~\ref{fig:f4}, Fig.~5 of Paper~I, and Fig.~1 of \cite{jose:98}
demonstrate a very good qualitative agreement, especially given that
they were obtained with different nuclear networks, reaction rates,
initial isotope abundances, and computer codes.

\begin{figure}
\includegraphics[width = 90mm]{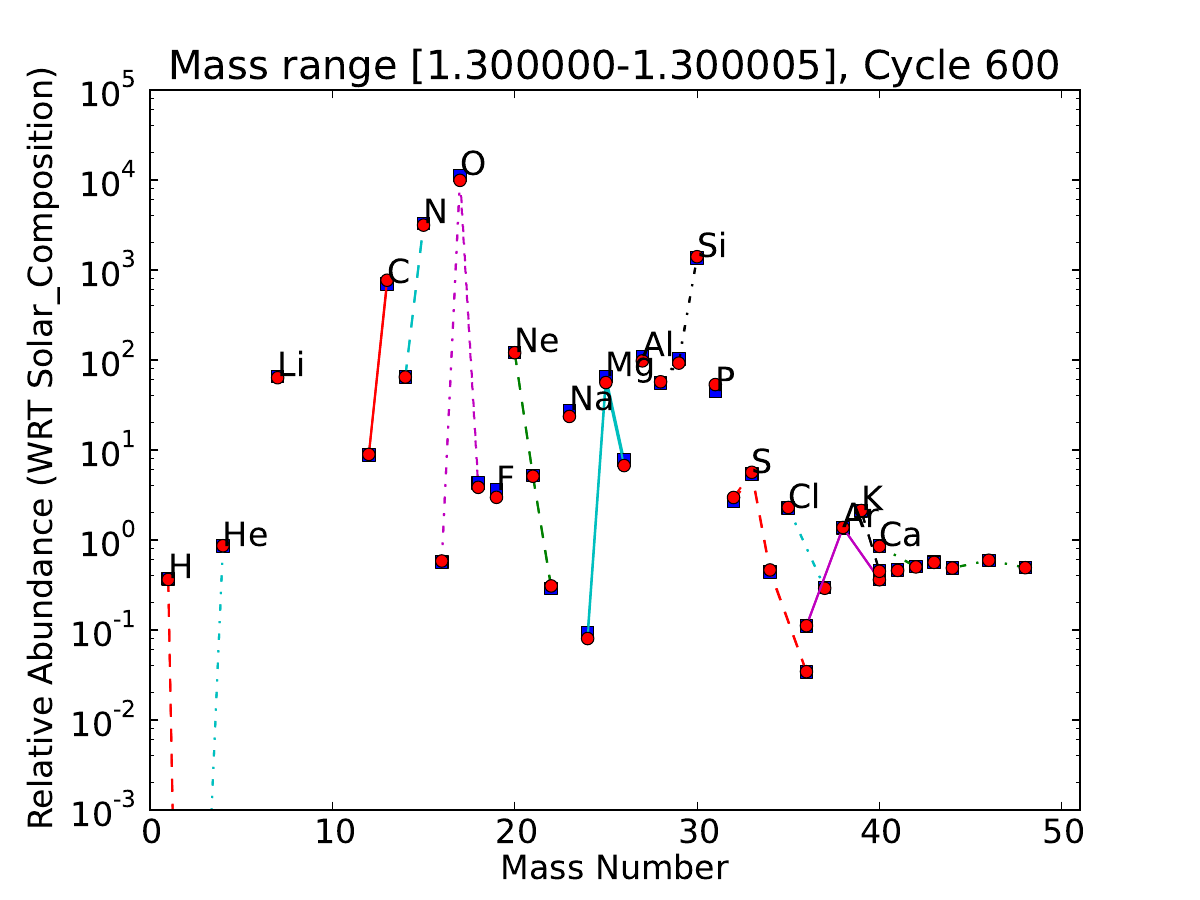}
\caption{Solar-scaled final abundances from two $1.3\,M_\odot$ ONe
         nova models calculated using the largest ({\tt nova.net},
         red circles) and intermediate
         ({\tt nova\_ONe.net}, blue squares, the latter being overlapped
         by the red circles) nuclear networks. For these models,
         $T_{\rm WD} =20$ MK, and $\dot{M} = 2\times 10^{-10}\,M_\odot/\mbox{yr}$.
         }
\label{fig:f7}
\end{figure}

\begin{figure}
\includegraphics[width = 90mm]{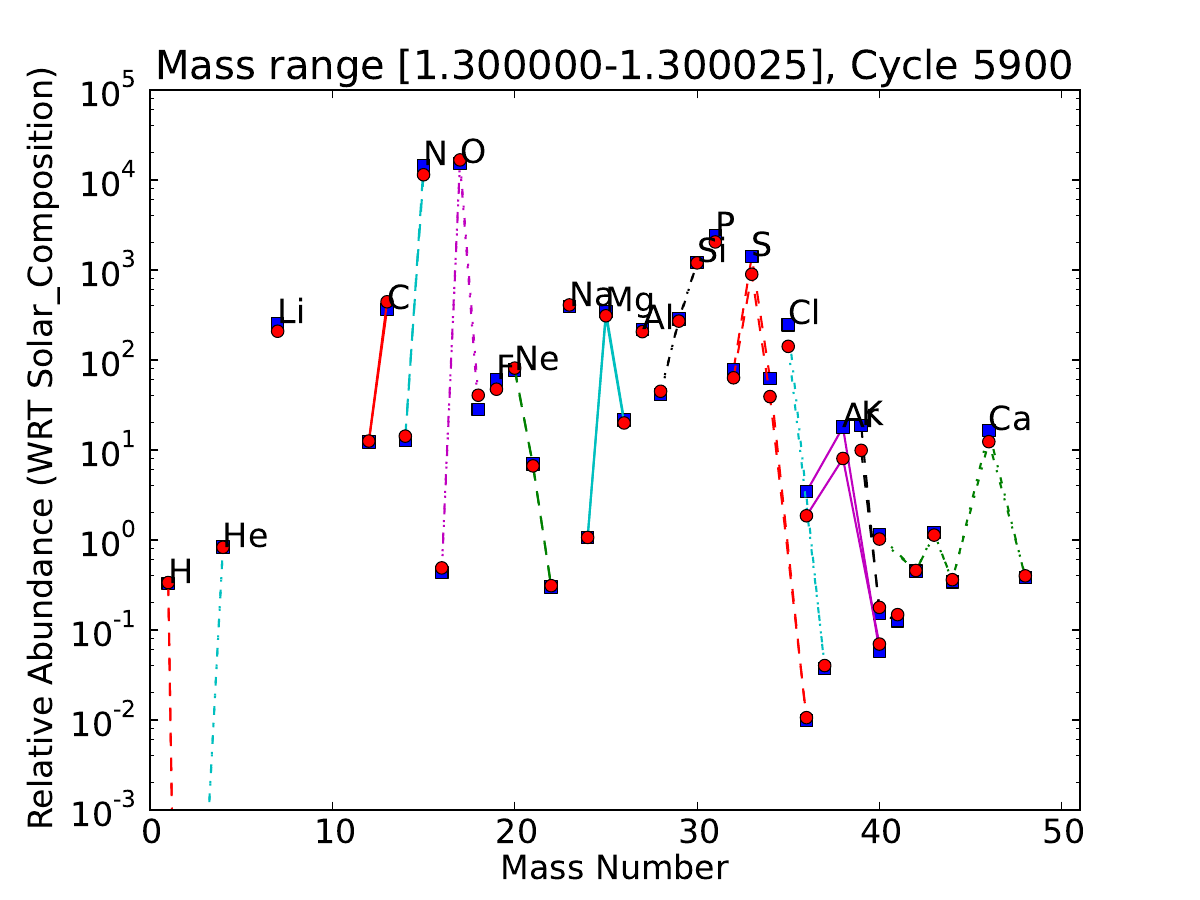}
\caption{Same as in Fig.~\ref{fig:f7}, but for $T_{\rm WD} = 7$ MK, and $\dot{M} = 10^{-11}\,M_\odot/\mbox{yr}$.
         $T_\mathrm{max}$-trajectories from these simulations are shown in Fig.~\ref{fig:f1}D.
         For such extreme cases leading to the highest values of $T_\mathrm{max}$, the switching to the
         {\tt nova.net} nuclear network is necessary.
         }
\label{fig:f8}
\end{figure}

\begin{figure}
\includegraphics[scale = 0.75, bb = 135 220 500 700]{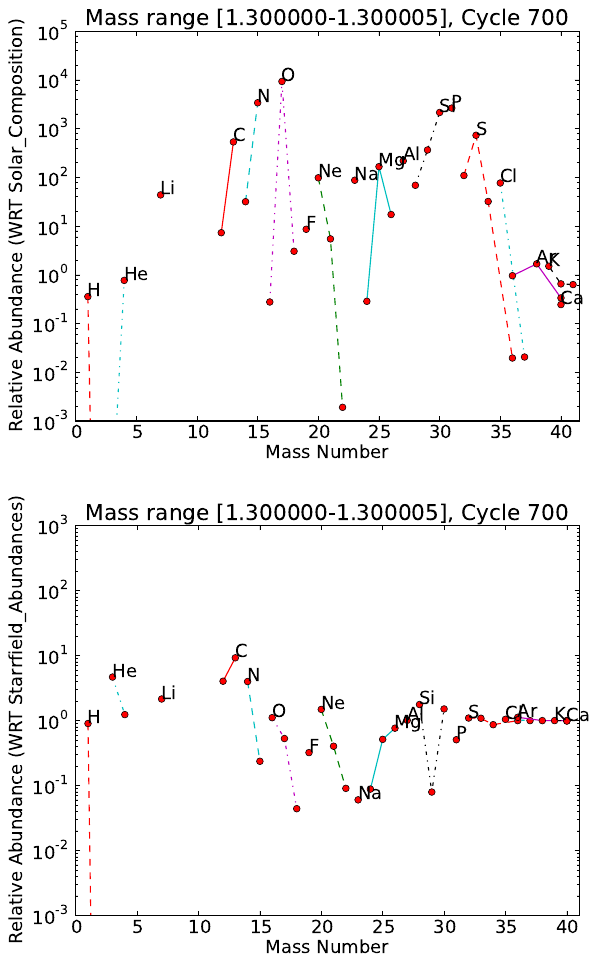}
\caption{Upper panel: solar-scaled final abundances in the expanding envelope of our $1.3\,M_\odot$ ONe nova model
         with $T_{\rm WD} = 12$ MK and $\dot{M} = 2\times 10^{-10}\,M_\odot/\mbox{yr}$.
         The initial abundances for these simulations have been taken from \protect\cite{politano:95}.
         Lower panel: the ratios of our final abundances to
         those reported by \protect\cite{starrfield:09} for a similar nova model with $M_{\rm WD} = 1.35\,M_\odot$,
         $T_{\rm WD} = 12$ MK, and $\dot{M} = 1.6\times 10^{-10}\,M_\odot/\mbox{yr}$ (I2005A in their Table~5),
         who used the same initial abundances.
         }
\label{fig:f9}
\end{figure}

\begin{figure}
\includegraphics[scale = 0.75, bb = 135 220 500 700]{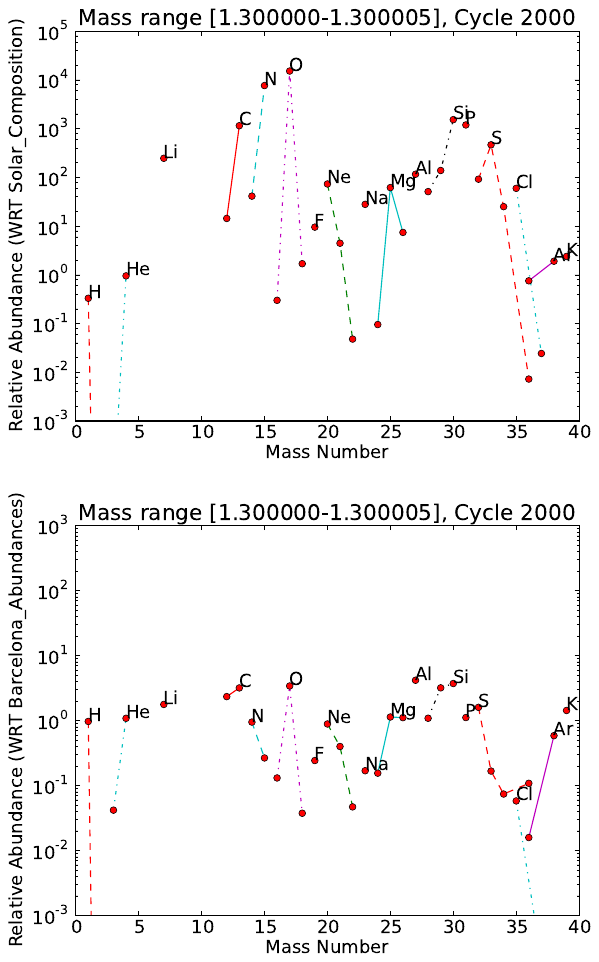}
\caption{Same as in Fig.~\ref{fig:f9}, but here we have used the Barcelona initial abundances and
         compare our MESA/NuGrid simulation results with those reported and plotted for a similar ONe nova model
         by \protect\cite{jose:98} in their Table~3 (the ONe6 column) and Fig.~3.
         }
\label{fig:f10}
\end{figure}

The final abundances from four nova models are shown in
Figs.~\ref{fig:f7} and \ref{fig:f8}. Two models are characterized by
high WD temperature and relatively fast accretion, calculated using two
different networks ({\tt nova.net} and {\tt nova\_ONe.net},
Fig.~\ref{fig:f7}). Their $T_\mathrm{max}$-trajectories (Fig.~\ref{fig:f1}C) and final
abundances (Fig.~\ref{fig:f7}) are really similar.
For the nova models with the coldest WDs and
accreting with the slowest rates, the
$T_\mathrm{max}$-trajectories are more different (Fig.~\ref{fig:f1}D), although the relative
nova yields do not show significant variations between the two cases (Fig.~\ref{fig:f8}).
In the upper panel of Fig.~\ref{fig:f9}, we plot our final MPPNP abundances
in the $1.3 M_\odot$ ONe nova model with $T_\mathrm{WD} = 12$ MK and
$\dot{M} = 2\times 10^{-10} M_\odot/\mathrm{yr}$. In this case, both the
MESA and MPPNP simulations have used the Politano composition for the
accreted material (the lower panel of Fig.~\ref{fig:f3}). In the lower
panel of Fig.~\ref{fig:f9}, we show the ratios of our abundances to the
corresponding final abundances from the ONe nova model I2005 that was
computed by \cite{starrfield:09} for the similar parameters, including
the same pre-mixed initial composition. Most of the abundances agree
with those of \cite{starrfield:09} within a factor of ten. A similar
comparison for our ONe nova model and its counterpart from
\cite{jose:98} in Fig.~\ref{fig:f10}, both computed with the Barcelona
initial composition (the upper panel of Fig.~\ref{fig:f3}), leads to the
same conclusion. The final abundances from the two independent works
with which we compare our results differ among each other as much as
each differs from our results. A more detailed comparison would require
simultaneous access to all participating simulation tools, and may not
be justified given the present accuracy of available observable
constraints.

\begin{figure}
\includegraphics[scale = 0.75, bb = 135 220 500 700]{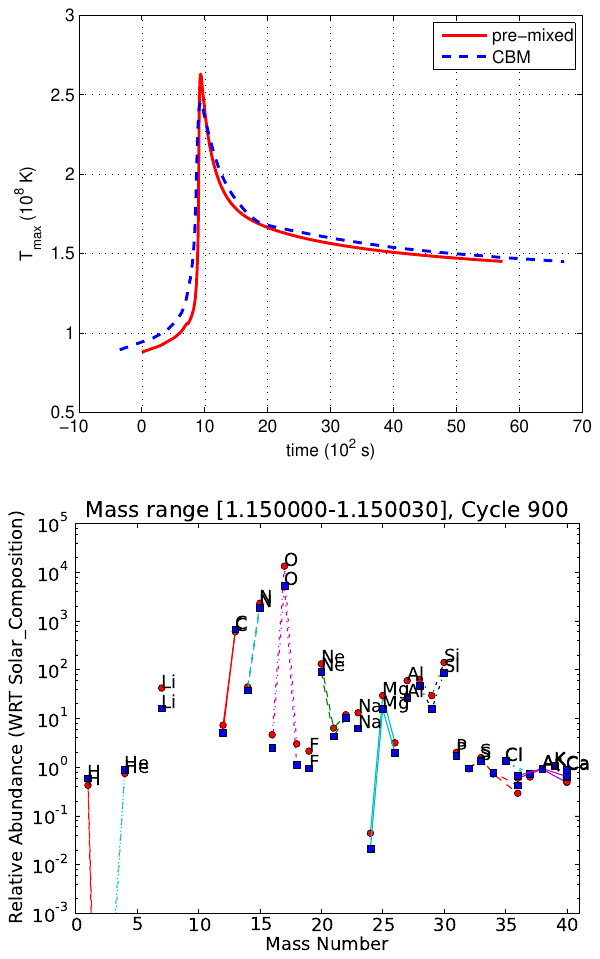}
\caption{A comparison of $T_{\rm max}$-trajectories (the upper panel) and final abundances (the lower panel) from two of our
         $1.15\,M_\odot$ ONe nova simulations with
         $T_{\rm WD} = 12$ MK and $\dot{M} = 2\times 10^{-10}\,M_\odot/\mbox{yr}$.
         The solid red curve and red circles are the results obtained with the 50\% pre-mixed
         initial abundances in the accreted envelope, while the dashed blue curve and blue squares represent the case
         with the solar-composition accreted material and convective boundary mixing (CBM) modeled using the same method
         and CBM parameter $f_{\rm nova} = 0.004$ that we used in the $1.2\,M_\odot$ CO nova model in Paper~I.
         }
\label{fig:f11}
\end{figure}

\begin{figure}
\includegraphics[scale = 0.75, bb = 135 220 500 700]{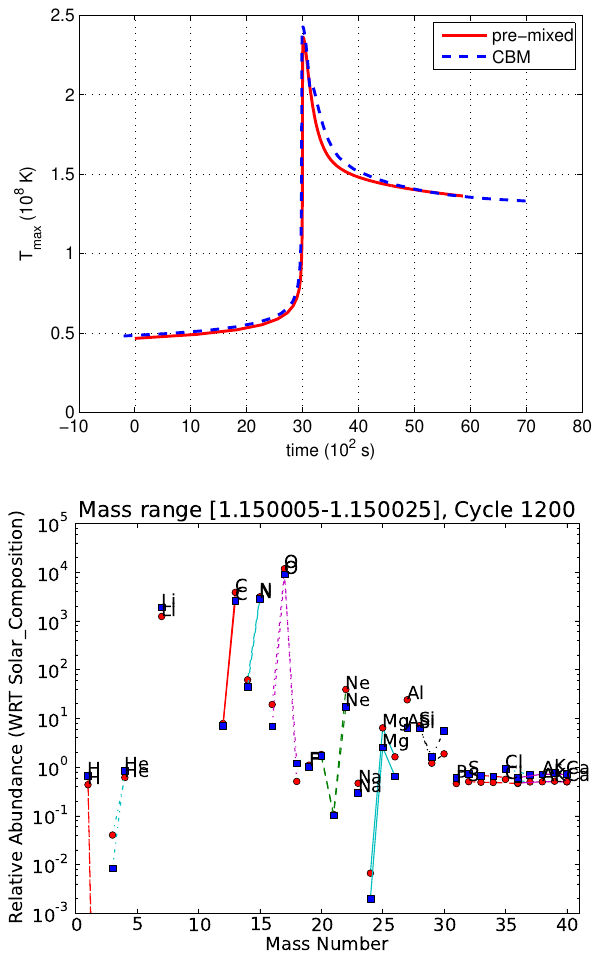}
\caption{Same as in Fig.~\ref{fig:f11}, but for the corresponding CO nova model.
         }
\label{fig:f12}
\end{figure}

Our MESA nova models with the
CBM have also been post-processed with the MPPNP code, and the resulting
final abundances are compared with those obtained for the corresponding
nova models without CBM but with the pre-mixed accreted envelopes in the
lower panels of Figs.~\ref{fig:f11} and \ref{fig:f12} for the ONe and CO
novae, respectively. The comparison between the final abundances shows a
very good agreement for both the ONe and CO nova models. This gives a
support to the widely-adopted 1D nova model in which the CBM is mimicked
by assuming that the accreted envelope has been pre-mixed with the WD's
material.

\begin{figure}
\includegraphics[scale = 0.75, bb = 135 220 500 700]{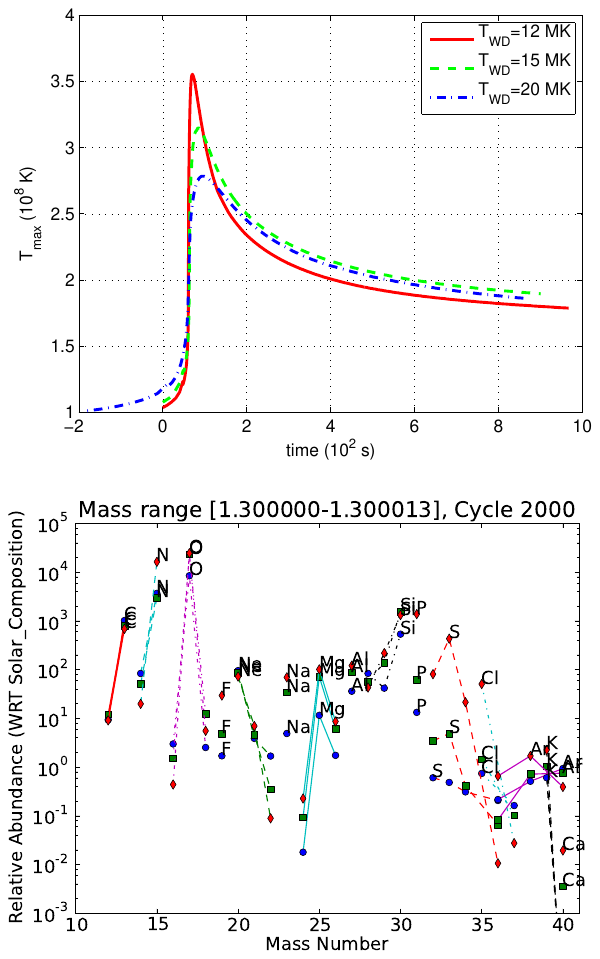}
\caption{A comparison of $T_{\rm max}$-trajectories (the upper panel) and solar-scaled final abundances (the lower panel)
         from three of our $1.3\,M_\odot$ pre-mixed ONe nova models computed for the same accretion rate,
         $\dot{M} = 2\times 10^{-10}\,M_\odot/\mbox{yr}$,
         but different WD initial central temperatures (the symbols in the lower panel have
         the same color coding as their corresponding
         curves in the upper panel). The initial isotope abundances are those of the Barcelona group.
         }
\label{fig:f13}
\end{figure}

\subsection{Single-Zone Computations: a Tool for Uncertainty and Sensitivity Studies}
\label{sec:ppn}

\begin{figure}
\centering
\includegraphics[scale = 0.75, bb = 135 220 500 700]{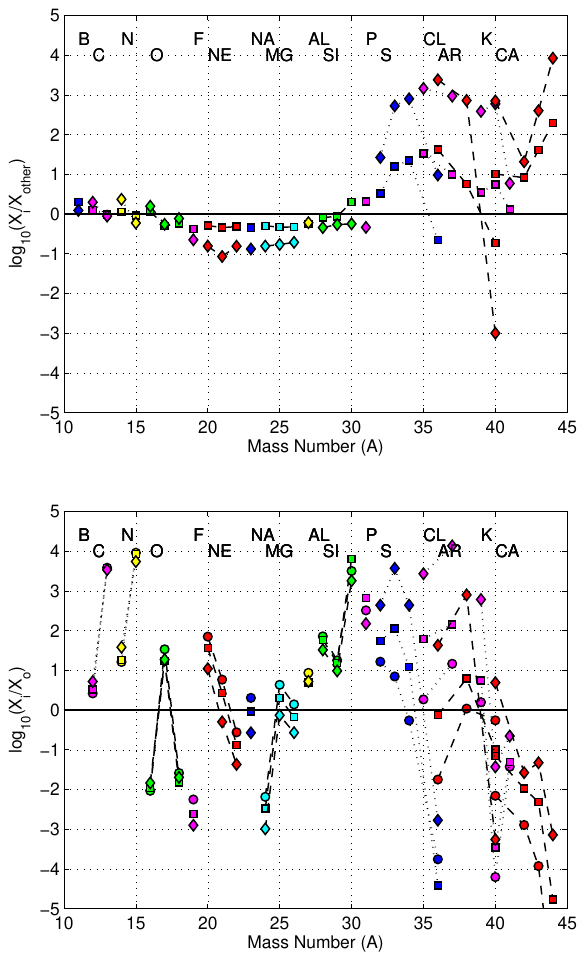}
\caption{Lower panel: the final solar-scaled abundances of stable isotopes calculated with the SPPN code
         using the trajectories from the upper panel of Fig.~\ref{fig:f13} that correspond to the three different
         WD's initial central temperatures: $T_\mathrm{WD} = 12$ MK (diamonds), 15 MK (squares), and 20 MK (circles).
         Upper panel: the abundances from the first two sets are plotted as ratios with respect to the abundances
         from the last set. The unstable isotopes were allowed to decay.
        }
\label{fig:f14}
\end{figure}

In this section, we discuss the post-processing simulations that use
$T_\mathrm{max}$-trajectories extracted from the MESA $1.3 M_\odot$
pre-mixed ONe nova models (shown in Fig.~\ref{fig:f13}) with the  mass
accretion rate $\dot{M} = 2\times 10^{-10} M_\odot/\mathrm{yr}$, but for
the three different WD initial central temperatures, $T_\mathrm{WD} =
12$, 15, and 20 MK . The accreted material had the Barcelona initial
composition. The $T_\mathrm{max}$ peak values for these models are 344,
313, and 267 MK, respectively (Table~\ref{tab:tab1}). The
post-processing nucleosynthesis calculations for these trajectories are
performed using the NuGrid SPPN code. The solver and nuclear physics
packages are the same that were used in our multi-zone post-processing
nova computations (\S \ref{sec:mppnp}), and we use the same nuclear
network with 147 isotopes (Fig.~\ref{fig:f5}). The final abundance
ratios for stable isotopes are presented in Fig.~\ref{fig:f14}. The
abundances were scaled using their corresponding solar values (the lower
panel) as well as the results obtained for the WD model with
$T_\mathrm{WD} = 20$ MK (the upper panel).

A comparison between the abundance distributions in Figs.~\ref{fig:f14}
and \ref{fig:f13} shows a qualitative agreement between the results from
the multi-zone complete models and from the single-zone
$T_\mathrm{max}$-trajectories. Indeed, the abundance patterns for
different groups of isotopes look similar. The main quantitative
difference is the more efficient production toward heavier species in
the second case, where O is more depleted, while Ar and K
isotopes\footnote{The Barcelona composition specifies the initial
abundances only for isotopes lighter than Ca.} are more efficiently
made. This difference is caused by the fact that in the multi-zone
post-processing simulations convective mixing reduces an ``average
temperature'' at which the nucleosynthesis occurs and it also constantly
replenishes the hydrogen fuel burnt at the base of the envelope by
bringing it from its outer parts, where $T\ll T_\mathrm{max}$. In
general, the results in Fig.~\ref{fig:f14} show how the increase of the
$T_\mathrm{max}$ peak value changes the relative abundance distribution.
We conclude that SPPN nucleosynthesis simulations with nova
$T_\mathrm{max}$-trajectories provide a proper qualitative indication of
the behavior of their corresponding complete MPPNP simulations. However,
they can be used only as a diagnostic for nova nucleosynthesis, e.g. in
reaction rate sensitivity studies.

The nucleosynthesis in ONe novae and its sensitivity to reaction rates
have been studied using post-processing \citep{iliadis:02} and full
hydrodynamic \citep{jose:07b,jose:10} models. As a result, three
reactions have been identified whose rate uncertainties have the most
significant impact on models' predictions: $^{18}$F(p,$\alpha)^{15}$O,
$^{25}$Al(p,$\gamma)^{26}$Si, and $^{30}$P(p,$\gamma)^{31}$S.

The $^{18}$F(p,$\alpha)^{15}$O reaction rate is crucial for
understanding the most intense 511 keV $\gamma$-ray emission that can be
observed from novae, which is produced by positron annihilation
associated with the decay of $^{18}$F. This nucleus is destroyed in a
nova environment via $^{18}$F(p,$\alpha)^{15}$O. The uncertainty in the
rate of this reaction therefore presents a limit to interpretation of
any future observed $\gamma$-ray flux. This reaction rate has been
evaluated recently in the compilation of \cite{iliadis:10}, where its
uncertainty was about a factor of 2 over the temperature range
characteristic of explosive hydrogen burning in novae. Since then,
several experiments \citep{beer:11,adekola:11,mountford:12} have been
performed to study the nuclear structure of the compound nucleus,
$^{19}$Ne, above the proton threshold. As a result, this rate has been
updated \citep{adekola:11}, and its uncertainty has been reduced to a
factor of 1.2 at 0.1 GK.

The $^{25}$Al(p,$\gamma)^{26}$Si reaction rate was evaluated for the
first time by \cite{iliadis:01}. Understanding this rate is important
for an accurate prediction of the yield of $^{26}$Al synthesized in ONe
novae. The isotopic abundance ratio of $^{26}$Al to $^{27}$Al serves as
a marker for identifying the candidate sources for presolar meteoritic
grains. Due to the lack of knowledge of the properties of the proton
resonances in $^{26}$Si, the uncertainty in this rate at 0.1 -- 0.4 GK
ranges over 4 orders of magnitude \citep{iliadis:02}. To reduce  the
uncertainty, the proton resonances in $^{26}$Si have been studied
experimentally, e.g. by \cite{matic:10}, \cite{chen:12}, and others,
and, as a result, the most recent evaluation of the
$^{25}$Al(p,$\gamma)^{26}$Si rate by Jun Chen (2010, private communication) reports an
uncertainty\footnote{Here and below, the uncertainty means that a real rate lies within
its recommended value multiplied and divided by the given factor.} 
of about a factor of 670 over the temperature range of novae.

Finally, the reaction $^{30}$P(p,$\gamma)^{31}$S drives the nuclear
activity in ONe novae in the atomic mass region above $A = 30$, and
affects the isotopic abundance ratio of $^{30}$Si to $^{28}$Si, which in
turn helps identify novae as a potential origin for some SiC presolar
grains. Its rate was first estimated by \cite{rauscher:00} via
Hauser-Feschbach statistical calculations, and was further discussed in
\cite{iliadis:01}. The properties of the proton resonances in $^{31}$S
were also largely unknown at the time, and therefore, this rate too
suffered from an uncertainty ranging over 4 orders of magnitude for nova
temperatures \citep{iliadis:02}. In recent years, the nuclear structure
of $^{31}$S has been studied extensively
\citep{ma:07,wrede:09,parikh:11,doherty:12}, and now the most updated
rate has an uncertainty of a factor of $\sim 17$ at 0.3 GK
\citep{parikh:11}.

\begin{figure}
\includegraphics[width = 84mm]{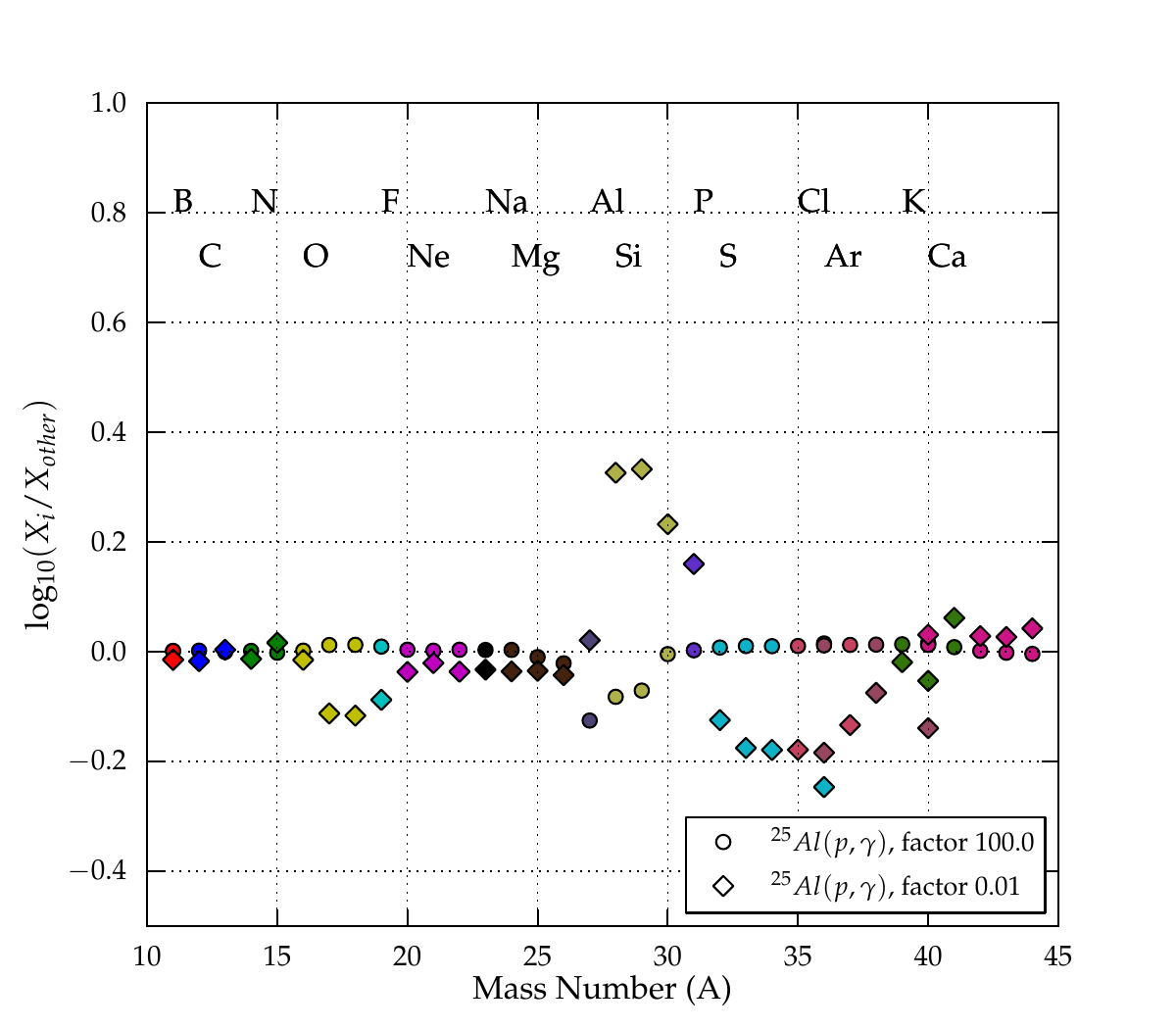}
\caption{Comparison of the final abundances of stable isotopes calculated with the SPPN code using the $T_\mathrm{WD} = 12$ MK
         trajectory from the upper panel of Fig.~\ref{fig:f13} for the recommended set of reaction rates ($X_\mathrm{other}$) with
         the abundances ($X_i$) obtained with the rate of the reaction $^{25}$Al(p,$\gamma)^{26}$Si increased (circles) and
         decreased (diamonds) by the factor of 100, like in the nuclear sensitivity study of \protect\cite{iliadis:02}.
        }
\label{fig:f15}
\end{figure}

\begin{figure}
\includegraphics[width=84mm]{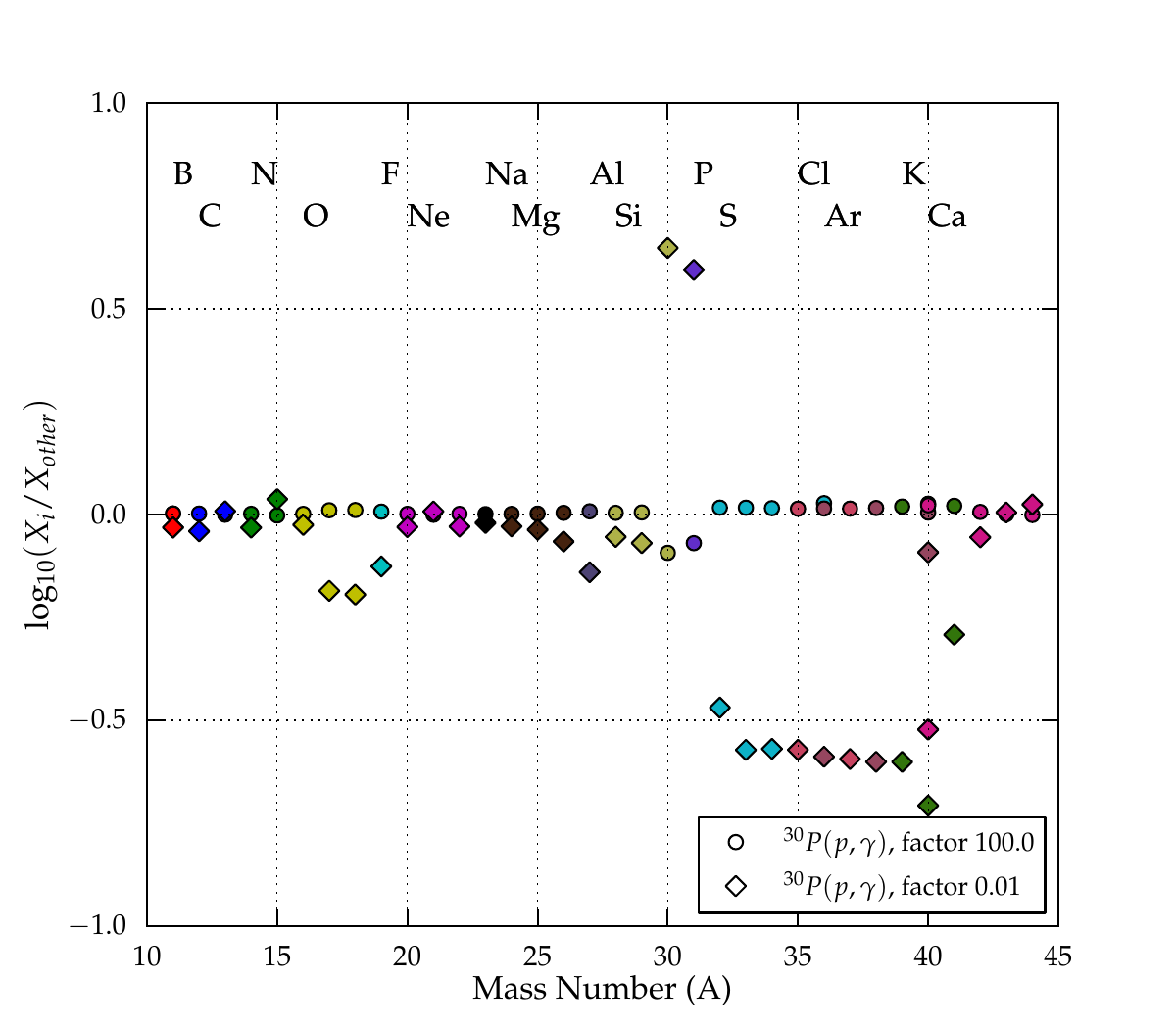}
\caption{Same as in Fig.~\ref{fig:f15}, but for the reaction $^{30}$P(p,$\gamma)^{31}$S.
        }
\label{fig:f16}
\end{figure}

We have calculated the effect of uncertainties in the rates of two of
the aforementioned nuclear reactions on final nova abundances with the
NuGrid SPPN code. In order to estimate the maximum effect, we have used
the $T_\mathrm{WD} = 12$ MK trajectory from the upper panel of
Fig.~\ref{fig:f13} that has the highest value of $T_\mathrm{max} = 344$
MK. Figs.~\ref{fig:f15} and \ref{fig:f16} show the final relative
abundances of stable isotopes resulting from varying the
$^{25}$Al(p,$\gamma)^{26}$Si and $^{30}$P(p,$\gamma)^{31}$S reaction
rates, respectively, within their (largest) 4 orders of magnitude
uncertainties used by \cite{iliadis:02}. Our results in
Fig.~\ref{fig:f16} are in good agreement with those presented by
\cite{iliadis:02} in their Fig.~3d. In particular, for both the
sensitivity studies the final abundances are more affected by reducing
the rates considered, starting respectively from $^{28}$Si and
$^{30}$Si.

The destruction of $^{25}$Al is mainly due to the
$^{25}$Al($\beta$$^+$)$^{25}$Mg and $^{25}$Al(p,$\gamma)^{26}$Si
reactions. Reducing the $^{25}$Al(p,$\gamma)^{26}$Si, the
nucleosynthesis path synthesizes less $^{26}$Si and $^{27}$P, therefore
reducing the flux through
this nucleosynthesis channel. On the other hand, $^{25}$Al is
accumulated causing a more efficient decay to $^{25}$Mg. Starting from
$^{25}$Mg, the proton capture chain
$^{25}$Mg(p,$\gamma)^{26}$Al(p,$\gamma)^{27}$Si(p,$\gamma)^{28}$P($\beta
$$^{+}$)$^{28}$Si causes a higher $^{28}$Si production. However, the
suppression of the $^{25}$Al(p,$\gamma)^{26}$Si channel makes the proton
capture nucleosynthesis overall slower, reducing the production in the
heavier S-Cl region and increasing the yields of Si and P in the final
nova ejecta.

The $^{30}$P(p,$\gamma)^{31}$S reaction is the dominant destruction channel for
$^{30}$P. Therefore, reducing its rate makes $^{30}$P a bottleneck for further
proton-capture nucleosynthesis, reducing the production of heavier species.
The isotope $^{30}$P is accumulated, and also $^{30}$Si
via $^{30}$P($\beta$$^{+}$)$^{30}$Si.
As a result, more $^{30}$Si and $^{31}$P are made, and less between S and $^{40}$Ca.

The differences between the final nova abundances caused by the variations of the
reaction rates are much smaller than those obtained for the slightly different
WD's initial central temperatures in Fig.~\ref{fig:f14}.
Therefore, these nuclear uncertainty studies for novae are particularly important
for detailed comparisons of the isotopic composition in a given atomic mass range, e.g.
from presolar grain measurements.

Note that these tests based on changing a single reaction rate are quite common in
nuclear astrophysics studies. This needs to be done carefully, since a consistent
uncertainty study for nuclear reaction rates of the nearby species
should also be included. It is not the purpose of this section to carry out such a detailed
analysis for $^{25}$Al(p,$\gamma)^{26}$Si and $^{30}$P(p,$\gamma)^{31}$S.
The point that we wanted to stress here is that single-zone trajectories can be used
as tools for exploratory studies to identify which reactions are relevant for the
nucleosynthesis (i.e., for sensitivity studies), and within which errors reaction
rates are needed to not affect stellar abundance predictions
(i.e., for uncertainty studies).

\section{Effects Caused by $^3$He Burning}
\label{sec:he3mix}

\begin{figure}
\includegraphics[scale = 0.6, bb = 120 220 500 560]{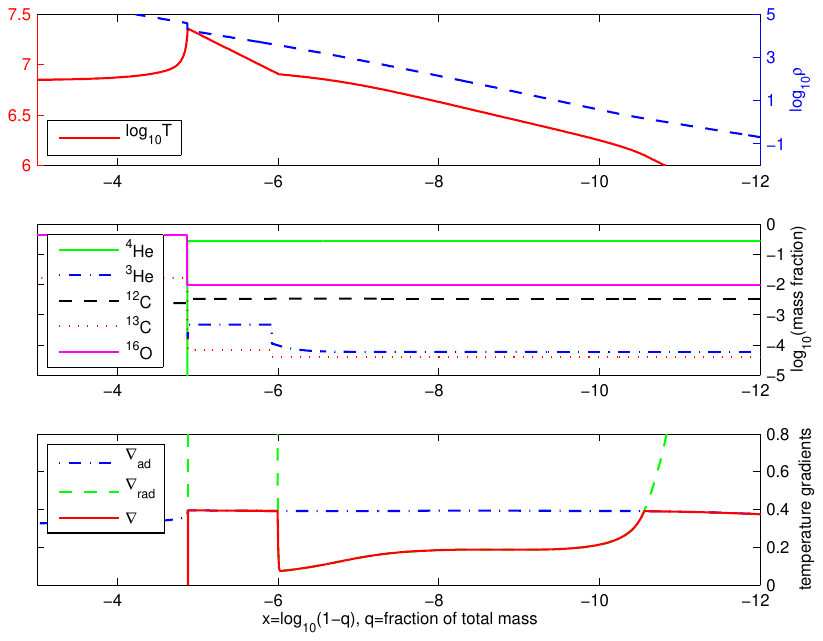}
\caption{A snapshot of profiles of relevant stellar structure parameters in the envelope of our $1.3\,M_\odot$ ONe nova model with
         $T_{\rm WD} = 7$ MK and $\dot{M} = 10^{-11}\,M_\odot/\mbox{yr}$ at the moment following the $^3$He ignition that
         has triggered convection. Like in the CO nova model with the cold WD and low accretion rate
         discussed in Paper~I, the low $T_{\rm WD}$ and $\dot{M}$ values favour the accumulation of $^3$He in a slope
         adjacent to the core-envelope interface followed by its ignition at a relatively low $T$,
         before the major thermonuclear runaway ensues.
        }
\label{fig:f17}
\end{figure}

\begin{figure}
\includegraphics[scale = 0.6, bb = 120 220 500 560]{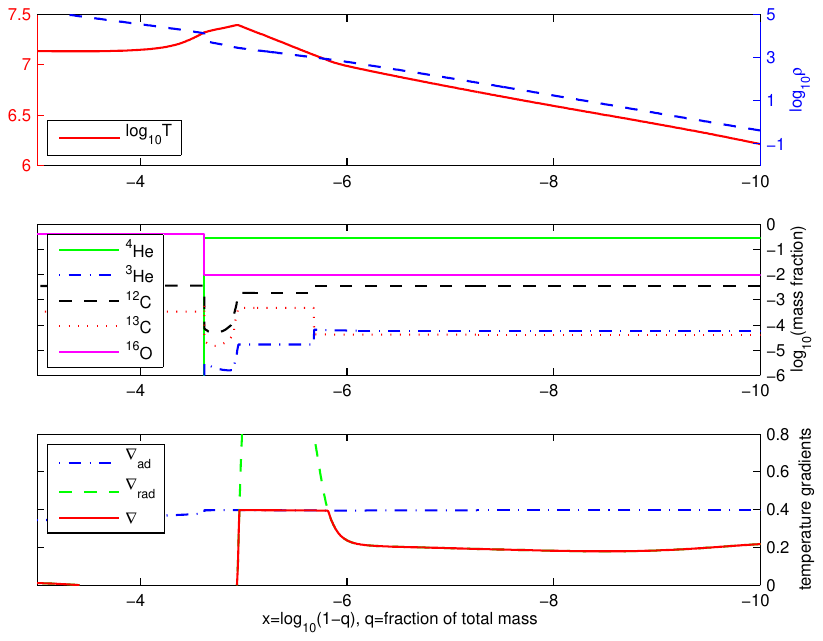}
\caption{A snapshot similar to that shown in Fig.~\ref{fig:f17}, but for our $1.15\,M_\odot$ ONe nova model with
         $T_{\rm WD} = 15$ MK and $\dot{M} = 10^{-10}\,M_\odot/\mbox{yr}$. The interplay between the $^3$He production
         and destruction shifts the peak temperature away from the core-envelope interface, located at the leftmost steps on
         the abundance profiles (the middle panel), to a place where the $^3$He
         burning generates the maximum energy. This leads to the formation of a convective zone separated from
         the interface by a buffer zone. Note that this model and the one in Fig.~\ref{fig:f17} accrete
         solar-composition material.
        }
\label{fig:f18}
\end{figure}

In Paper~I, we have found, for the first time, that in the extreme case
of a very cold CO WD, e.g. with $T_\mathrm{WD} = 7$ MK, accreting
solar-composition material with a very low rate, say $\dot{M} = 10^{-11}
M_\odot/\mathrm{yr}$, the incomplete pp~I chain reactions lead to the
{\em in situ} synthesis of $^3$He in a slope adjacent to the base of the
accreted envelope and that the ignition of this $^3$He triggers
convection before the major nova outburst. The $^3$He burning continues
at a relatively low temperature, $T\approx 30$ MK, approximately until
its abundance is reduced below the solar value, only after that the
major nova outburst ensues triggered by the reaction
$^{12}$C(p,$\gamma)^{13}$N. Although this is an interesting variation of
the nova scenario,\footnote{It is interesting that $^3$He was considered
as the most likely isotope to trigger a nova outburst by
\cite{schatzman:51}.} because the CO enrichment of the accreted envelope
is produced by the $^3$He-driven CBM in this case, its observational
frequency is expected to be very low (Paper~I). Here, we confirm this
result for ONe novae (Fig.~\ref{fig:f17}). Furthermore, we have found
that in the $1.15\,M_\odot$ ONe WD with a relatively high initial
central temperature, $T_{\rm WD} = 15$ MK, accreting the
solar-composition material with the intermediate rate, $\dot{M} =
10^{-10}\,M_\odot/\mbox{yr}$, the interplay between the $^3$He
production and destruction in the vicinity of the base of the accreted
envelope first leads to a shift of $T_\mathrm{max}$ away from the
core-envelope interface followed by the formation of a thick radiative
buffer zone that separates the bottom of the convective envelope from
the WD surface (Fig.~\ref{fig:f18}). This result is obtained only for
the case when an ONe WD accretes solar-composition material and it
almost disappears in the models with the pre-mixed accreted envelopes.
Given that the formation of the radiative buffer zone is revealed in the
computations with the more realistic nova model parameters, such cases
can probably be observed. We defer a more detailed study of this
peculiar case and its possible consequences to a future work. In
particular, it would be interesting to see if the CBM can cross the
buffer zone and reach the WD core.

\section{Comparison With Observed Chemical Compositions of Novae}
\label{sec:observations}

\subsection{Element Abundances From Optical, Ultraviolet and Infrared Spectroscopy}
\label{subsec:spectroscopy}

\begin{figure}
\includegraphics[scale = 0.75, bb = 135 170 500 660]{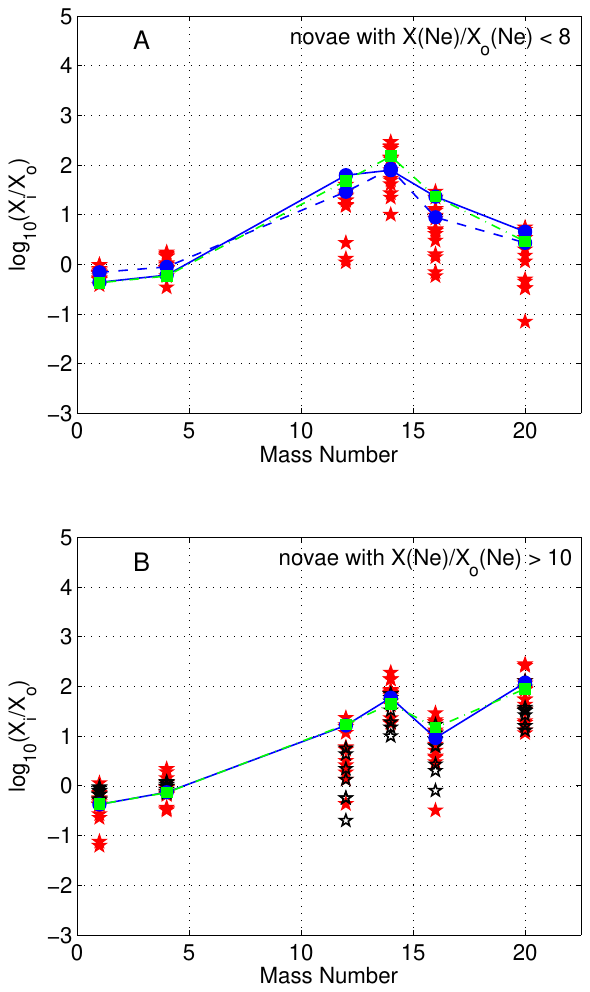}
\caption{Comparison of the H, He, C, N, O, and Ne mass fractions in novae from optical, ultraviolet and infrared spectroscopy
         (the filled red and open black star symbols are data from \protect\citealt{gehrz:98} and \protect\citealt{downen:13}, 
         respectively) with the theoretical ones obtained in nova simulations.
         Blue circles connected by solid blue curves in panels A and B correspond to our CO and ONe
         nova models with 50\% pre-mixed accreted envelopes from Figs.~\ref{fig:f12} and \ref{fig:f11}. The dashed blue curve
         in panel A represents our results for a 25\% pre-mixed CO nova model, while green squares
         connected by dot-dashed green curves in panels A and B are data for nova models,
         CO5 and ONe3, from Tables 4 and 3 of \protect\cite{jose:98}.
        }
\label{fig:f19}
\end{figure}

Difficulties with determining element abundances from the spectra of
nova ejecta are discussed by \cite{gehrz:98}. The same authors have
compiled the largest set of mass fractions of H, He, and heavy elements,
such as C, N, O, Ne, in novae from optical and ultraviolet spectroscopy
that were published in the period from 1978 until 1997. In
Fig.~\ref{fig:f19}, we compare these abundances (the filled red star symbols)
as well as ONe nova abundances from the recent compilation by \citet[][the open black star symbols]{downen:13},
that include infrared spectroscopy data,
with the corresponding nucleosynthesis yields predicted by our $1.15 M_\odot$ pre-mixed nova models from
Figs.~\ref{fig:f11} and \ref{fig:f12}. We also include the abundance
predictions by \cite{jose:98}. For each of the novae listed in Table~2
of \cite{gehrz:98}, we have used all the available abundance
measurements, therefore some of the novae are represented by up to three
points for the same element in our Fig.~\ref{fig:f19} that correspond to
different data sources. We have divided the observed objects
into CO (panel A) and ONe (panel B) novae using the neon mass fraction ratios
$X(\mathrm{Ne})/X_\odot (\mathrm{Ne}) < 8$ and $X(\mathrm{Ne})/X_\odot (\mathrm{Ne}) > 10$ for
the first and second group. Neither panel shows a good agreement between the observed and predicted element
abundances for the novae, because both show the presence of large numbers of
observed objects with too low C and O abundances as compared to the
model predictions. The latter rather represent upper limits for the
observed data. We do not know how to interpret this discrepancy, other
than to assume that the observed material has been mixed with the
solar-composition material from the WD's companion. We have tried to
reduce the amount of CO WD's material in the pre-mixed accreted envelope
from 50\% to 25\% but this does not help much (compare the solid and
dashed blue curves in the panel A). Note that our model predictions
almost coincide with those of the Barcelona group in Fig.~\ref{fig:f19}.

\subsection{Presolar Grains}
\label{subsec:presolar_grains}

Absorption features due to carbide grains have been observed around
novae \citep[e.g.,][]{starrfield:97,gehrz:98}. This opens a possibility
to find in the solar system, hidden in carbonaceous pristine meteorites,
presolar grains that condensed around nova objects shortly before the
Sun formed, and possibly carrying the Ne-E component observed in
meteoritic samples \citep[][]{starrfield:97}. This scenario is confirmed
by the analysis of abundances for a small fraction of presolar grains
identified so far, which show distinctive isotopic signatures that may
be associated with nova nucleosynthesis: a minor fraction of the
presolar silicon carbides \citep[SiC
nova,][]{amari:01,amari:02,jose:07b,heck:07}, a few graphite grains
\citep[e.g.,][]{amari:01,zinner:07}, and oxides
\citep[e.g.,][]{gyngard:10,nittler:10,gyngard:11}. Oxide grains should
form in CO novae, whereas carbide grains around ONe novae, since the
ejected material shows the C to O abundance ratios less than unity in
the first case, and being C-rich in the last case (e.g.,
\citealt{amari:01}; also, see our Table \ref{tab:tab2}). However,
observational features of carbide dust have been observed also around CO
novae, and therefore this simple distinction cannot be made
\citep[][]{gehrz:02}. The distinctive isotopic signatures of nova grains
include extremely high excesses of $^{13}$C, $^{15}$N, overabundances of
$^{26}$Mg and $^{22}$Ne due to the later radiogenic decay of the
radioactive $^{26}$Al and $^{22}$Na (Table \ref{tab:tab2}), and high
$^{17}$O/$^{16}$O isotopic ratios, specifically for oxide grains
\citep[][]{clayton:76,amari:01,amari:02,jose:07b}. \cite{nittler:05}
suggested that, at least, part of the nova SiC grains are instead made
from material ejected by core-collapse SNe, showing typical nova
signatures coupled with $^{28}$Si and $^{44}$Ti excesses that cannot be
obtained from novae. 

\begin{table*}
\caption{Comparison of Selected Abundances in Pre-Mixed Nova Models$^{a}$}
\label{tab:tab2}
\begin{tabular}{ccccccc}
\hline
Abundance & Fig.~\ref{fig:f1} & CO5 & Fig.~\ref{fig:f13} & ONe3 & Fig.~\ref{fig:f12} & ONe6 \\
\hline
$\mathrm{C}/\mathrm{O}$ & 0.56 & 0.78 & 0.36 & 0.43 & 2.2 & 1.2 \\
$X_{22}$ & $6.1\times 10^{-7}$ & $2.9\times 10^{-7}$ & $2.6\times 10^{-4}$ & $5.3\times 10^{-5}$ & $1.1\times 10^{-3}$ &
$6.0\times 10^{-4}$ \\
$X_{26}$ & $2.1\times 10^{-4}$ & $4.7\times 10^{-5}$ & $3.1\times 10^{-4}$ & $9.3\times 10^{-4}$ & $8.4\times 10^{-4}$ &
$7.2\times 10^{-4}$ \\
\hline   
\end{tabular}

\medskip

$^{a}$The C to O number ratios and mass fractions of $^{22}$Na and $^{26}$Al are taken from our models
shown in the indicated figures and from Tables 3 and 4 of \cite{jose:98} for their corresponding counterparts that
are denoted by CO5, ONe3, and ONe6.

\end{table*}

\begin{figure}
\includegraphics[scale = 0.4, bb = 10 120 500 600]{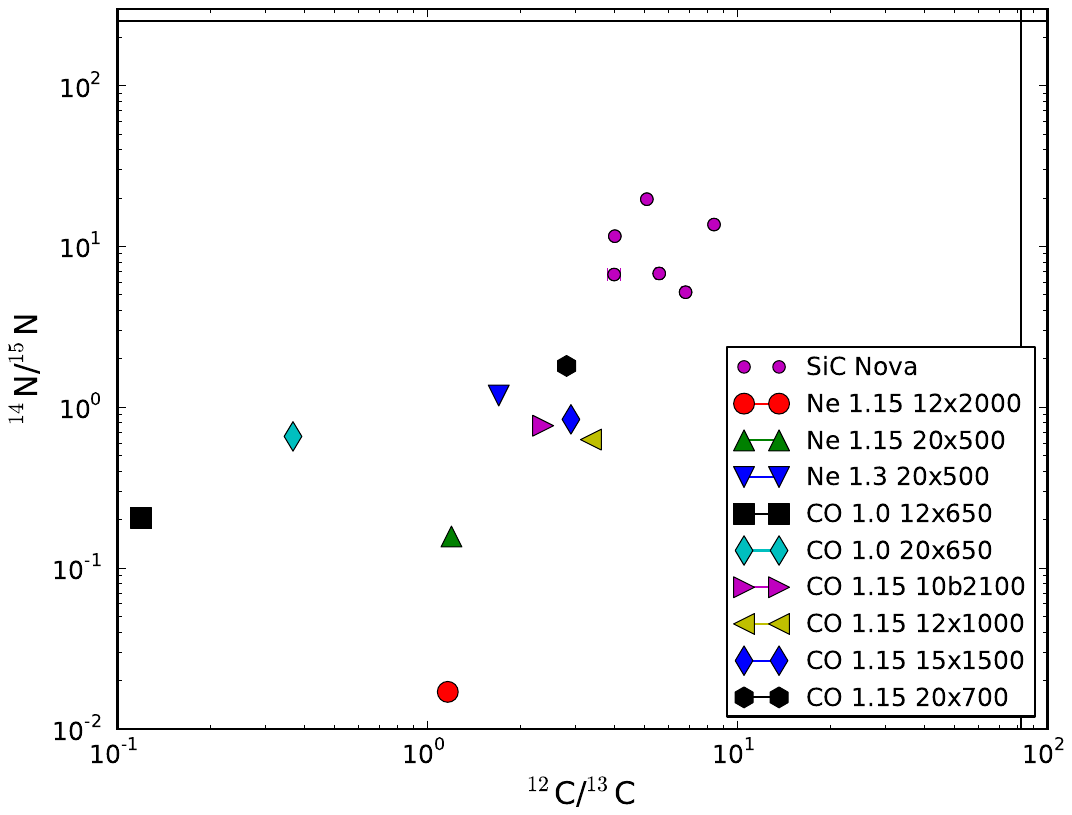}
\caption{Comparison of the C and N isotopic ratios from our ONe and CO nova models
         with presolar grain data. Black lines indicate the solar system values.
         Presolar grain data are from \protect\citet{amari:01,gao:97,hoppe:96,nittler:05}.
         Nova models are denoted by their composition (Ne or CO), WD mass ($M_\odot$) and initial central temperature (MK), 
         followed by a number of models in the evolutionary sequence. Their expanding envelopes are mixed as described in the text.
        }
\label{fig:f20}
\end{figure}

\begin{figure}
\includegraphics[scale = 0.4, bb = 10 120 500 600]{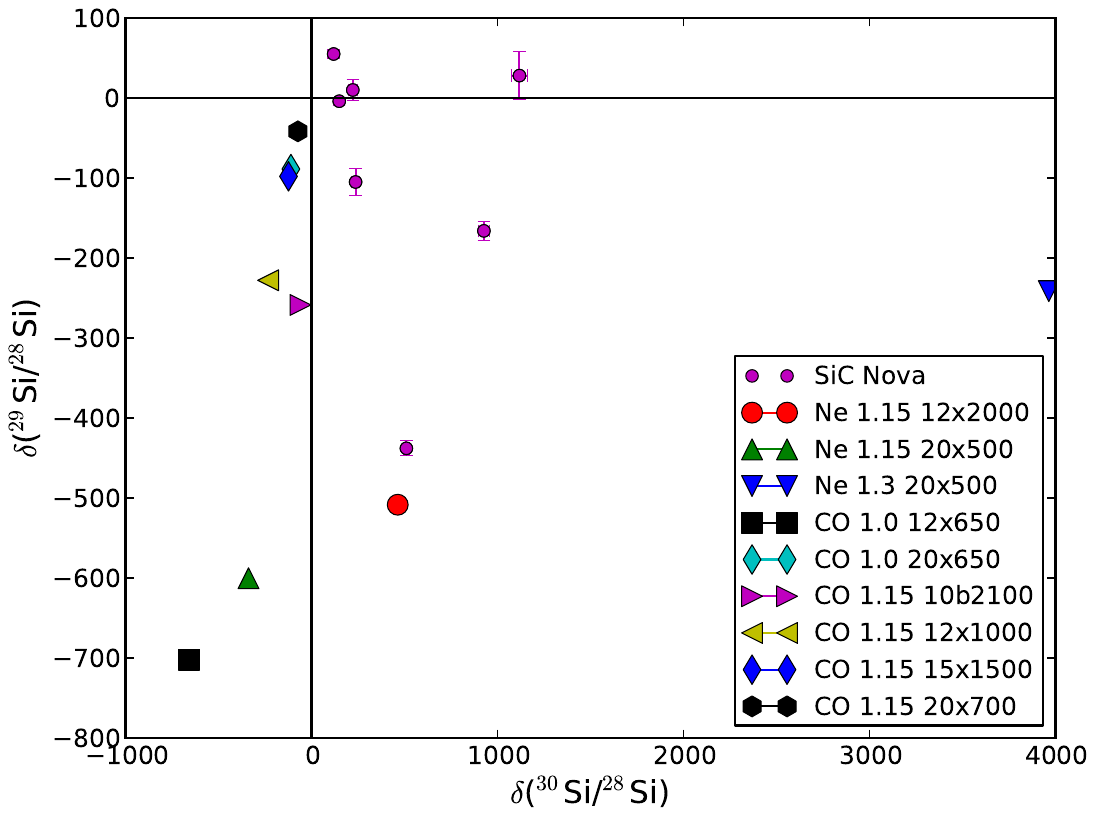}
\caption{Comparison of the Si isotopic ratios from ONe and CO nova models with single
         presolar grain data. For the isotopic ratios, the delta notations are used
         ($\delta$(ratio) = (stellar ratio/solar ratio - 1)$\times$1000).
         Black lines indicate the solar system values. The presolar grain data are from
         \protect\citet{amari:01,gao:97,hoppe:10,nittler:03,nittler:05}.
         Nova models are denoted by their composition (Ne or CO), WD mass ($M_\odot$) and initial central temperature (MK), 
         followed by a number of models in the evolutionary sequence. Their expanding envelopes are mixed as described in the text.
         }
\label{fig:f21}
\end{figure}

In this section, we compare the model results presented in the previous
sections for ONe novae with presolar SiC grains. Also plotted are the
results for CO novae from Paper I. We derive a single data point for the
isotope abundances from each nova model, using the mass average of all
zones in the last time step of post-processing nucleosynthesis
simulations, and use these data to compare stellar model
predictions with isotopic ratios measured in presolar grains. 
We assume that all the zones in the expanding envelope will eventually be ejected
after its radius has reached a few solar radii.
We do not distinguish between carbon and oxygen rich layers in the expanding nova
envelope. This results in an average isotopic composition for each
individual nova simulation and does not necessarily represent what
presolar grain incorporate.
A more detailed comparison between grains of possible nova origin and
ejected layers with nova models would require a dedicated investigation.
Figs.\,\ref{fig:f20} and \ref{fig:f21} compare presolar SiC nova grains
with our model results. For presolar grains, we used the Washington
University (St. Louis) database for data extraction \citep{hynes:09}.
The data mostly fall between the model and average solar system isotopic
ratios. Therefore, present models reproduce the data well, if some
contamination with solar system material is taken into account. Sample
contamination either happened on the asteroid parent body or in the lab.
Since no major and minor element concentrations were measured in the
grains, a mixing calculation cannot be performed here.

\begin{figure}
\includegraphics[width=84mm]{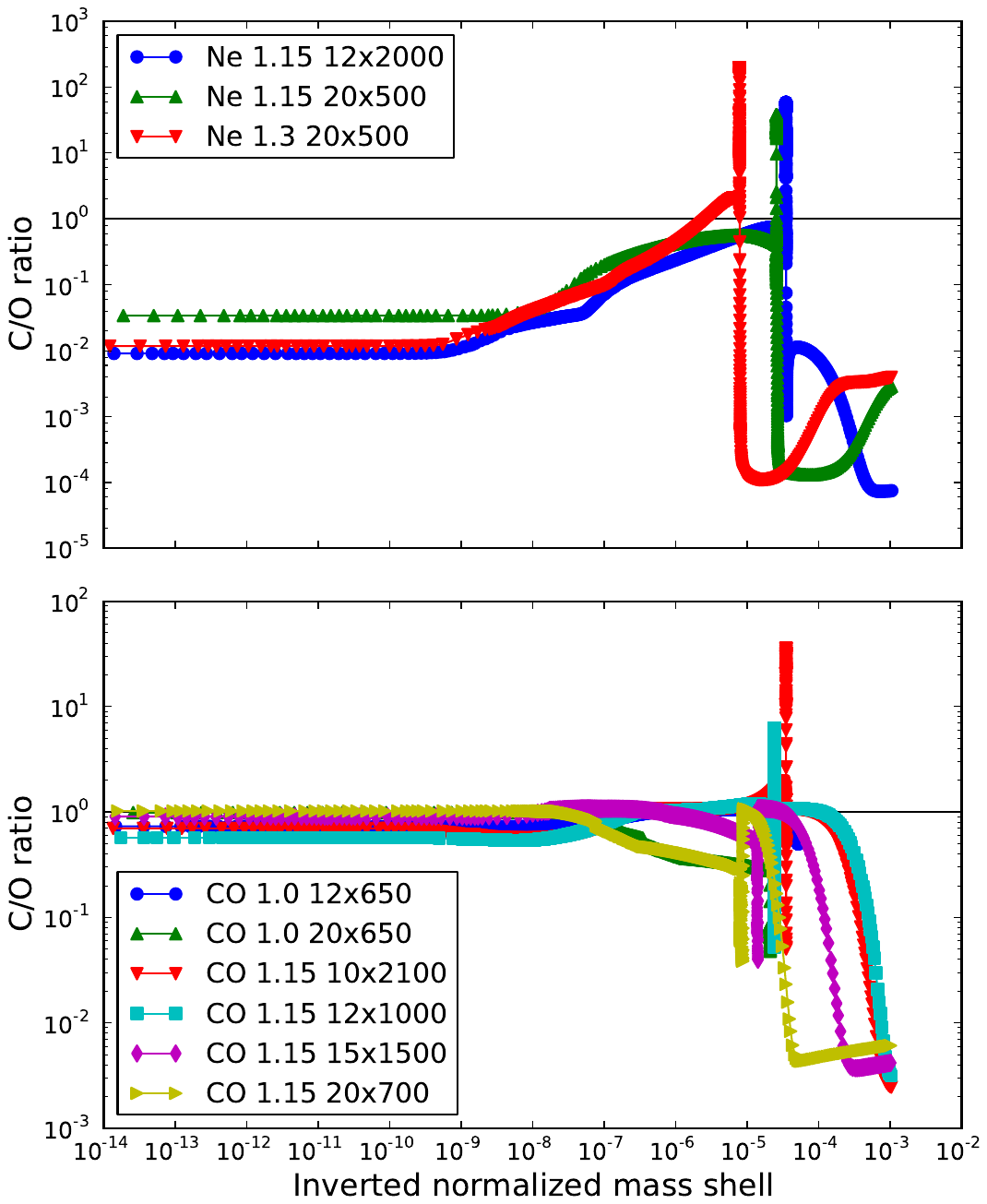}
\caption{C/O ratios in expanding envelopes of all nova models in the last time step. The C/O = 1 is indicated by 
         a straight line in each plot. 
         To plot all models on the same mass scale and point out differences in the outer layers better, the x-axis shows 
         the mass-shell for each model normalized from 0 to 1 and inverted. This means that the outermost shell lies 
         on the left side, while the innermost shell is on the right. The figure clearly demonstrates that mixing of 
         the whole envelope is not {\em a priori} valid, rather carbon rich zones have to be mixed and analyzed in order to obtain 
         a valid comparison with presolar SiC grains.}
\label{fig:f22}
\end{figure}

Equilibrium condensation calculations show that SiC grains are only expected
to condense in environments where the C/O ratio is larger than one
\citep[see, e.g.,][]{zinner:03}. Therefore, unless mixing of the ejecta takes
place, the chosen approach to mix all shells together, is not completely valid.
Fig.~\ref{fig:f22} shows the evolution of the C/O elemental ratio
in the ONe nova models (upper panel) and CO nova models from Paper I (lower panel).
To make changes on the surface visible and plot all models in one graph,
the mass layers were normalized from zero to one and inverted,
such that the outermost shell plots on the left side.
Only part of the nova ejecta is C-rich.
Without a clear mixing prescription for the ejecta -- if existing --
a definite comparison with presolar grains requires a careful study where
the analysis is performed grain-by-grain, and considering the isotopic signature
in different stellar zones.

In summary, we find a general agreement between our nucleosynthesis calculation trends and isotopic abundance
measurements from SiC grains with a possible nova origin. The comparison however only gives a trend of
our nova models since mixing in the ejecta and contamination with solar system material have to be further taken
into account and discussed. A more detailed grain-by-grain analysis is required for a consistent and constraining analysis,
including analyses of more isotopic constraints where available.

\section{Conclusion}
\label{sec:concl}

We have created the Nova Framework that allows to simulate the accretion of H-rich material
onto a white dwarf (WD) leading, for a suitable set of initial parameters, to a nova outburst,
and to post-process its accompanying nucleosynthesis.
The Nova Framework combines the state-of-the-art stellar evolution code MESA and post-processing nucleosynthesis
tools of NuGrid. It includes a number of CO and ONe WD models with different
masses and central temperatures (luminosities) that can be used in simulations of
nova outbursts. The use of the Nova Framework is facilitated by a number of shell scripts
that carry out the routine job necessary to coordinate the operation of the MESA and NuGrid codes.

Within the Nova Framework effort, we provide a large set of nucleosynthesis
calculations for nearly 50 CO and ONe nova models at the solar
metallicity ($Z = 0.02$) so far. To verify our calculations, we compare their results with
those published in the literature for similar nova models. The comparison shows
a very good qualitative agreement analogous to the case when different nova yields from the literature are compared
with each other.
Typical features of abundance signatures of novae from previous studies and observations
are confirmed. For instance, the large production factors for $^{13}$C, $^{15}$N, and $^{17}$O in
both CO and ONe novae, the synthesis of $^7$Li in CO novae, the overabundance of Ne as a distinctive
feature of ONe novae, as well as the accumulation of relatively large mass fractions of the radioactive
isotopes $^{22}$Na and $^{26}$Al in ONe novae (unfortunately, not detected by $\gamma$ telescopes yet).

The nucleosynthesis in novae is studied using the post-processing method.
We show that such a technique provides abundance predictions consistent with
stellar model calculations.
We study the impact on the final abundances of model parameters relevant
for novae, such as the mass, composition and initial central temperature of the underlying WD, and the mass accretion rate.
In particular, decreasing the initial central temperature causes
the outburst peak temperature to increase, which in turn enhances the production of isotopes in the region between
Si and Ca.
The main effect of the decrease of the accretion rate is an increase of the envelope mass $M_\mathrm{acc}$ accreted before
a nova outburst. The higher $M_\mathrm{acc}$ value results in a stronger explosion with
a higher peak temperature. For accretion of solar-composition material with a very low rate, 
e.g. $\dot{M} = 10^{-11} M_\odot/\mathrm{yr}$, onto a very cold WD, e.g. with $T_\mathrm{WD} = 7$ MK, 
we find the {\em in situ} synthesis of $^3$He
taking place via the incomplete pp I chain near the bottom of the accreted envelope. The ignition and burning of this $^3$He
at a relatively low temperature ($T\sim 30$ MK) triggers convection in the envelope before the major nova outburst, the latter
ensuing only after the $^3$He abundance has been reduced below its solar value. We demonstrate that
this happens not only in CO novae (Paper I) but also in ONe novae. Moreover, we reveal that
the interplay between the $^3$He production and destruction in the solar-composition envelope accreted with 
an intermediate rate, e.g.  $\dot{M} = 10^{-10}\,M_\odot/\mbox{yr}$, by the $1.15\,M_\odot$ ONe WD with 
a relatively high initial central temperature, e.g. $T_{\rm WD} = 15\times 10^6$ K,
leads to the formation of a thick radiative buffer zone that separates the bottom of the convective envelope from the WD surface.

We use three $T_\mathrm{max}$-trajectories, extracted from full nova models, in single-zone post-processing
nucleosynthesis computations with the NuGrid SPPN code.
We show that the results obtained in the one-zone simulations are qualitatively consistent with the yields from
complete models post-processed with the NuGrid multi-zone code MPPNP, including the impact of stellar parameters, like the
WD initial central temperature. Therefore, simple trajectories can be used as important
diagnostic tools for nova nucleosynthesis, e.g., for nuclear
sensitivity and uncertainty studies.
As example, we have studied the impact on nova nucleosynthesis of the
$^{25}$Al(p,$\gamma)^{26}$Si and $^{30}$P(p,$\gamma)^{31}$S reaction rates.
In particular, we find that abundance predictions are more affected by reducing those rates.

The interested reader can try to post-process our nova $T_\mathrm{max}$-trajectories with the NuGrid SPPN code
using our NuGrid Ubuntu ppn-nova virtual machine. 
Instructions on how to use this virtual machine can be found on {\tt http://nugridstars.org} 
in the subsection ``Virtual Box releases'' of the section ``Releases and Software Downloads''.

\section*{Acknowledgments} 

This research has been supported by the National
Science Foundation under grants PHY 11-25915 and AST 11-09174. This
project was also supported by JINA (NSF grant PHY 08-22648) and
TRIUMF. Falk Herwig acknowledges funding from Natural Sciences and Engineering Research Council of
Canada (NSERC) through a Discovery
Grant. MP also thanks the support from Ambizione grant of the SNSF
(Switzerland), and from EuroGENESIS.
Kiana Setoodehnia acknowledges an NSERC support through Alan Chen's Discovery Grant.
Reto Trappitsch is supported by NASA Headquarters under the NASA Earth
and Space Science Fellowship Program - Grant NNX12AL85H.

\bibliography{paper}

\label{lastpage}

\end{document}